\documentclass[conference]{IEEEtran}
\IEEEoverridecommandlockouts
\usepackage{amsmath,amssymb,amsfonts}
\usepackage{subcaption}
\usepackage{tikz}
\usepackage[ruled,vlined]{algorithm2e}
\usepackage{booktabs}
\usepackage{array}
\usepackage{graphicx}
\usepackage{physics}
\usepackage{cite}
\usepackage{hyperref}
\usepackage[nameinlink,capitalize]{cleveref} 
\usepackage{float}
\usepackage{textcomp}
\usepackage{xcolor}
\usepackage{tikz,pgfplots,pgfplotstable}
\pgfplotsset{compat=newest}
\usepackage{tikzit}
\tikzstyle{new style 1}=[circle, fill=white, draw=black, scale=0.7, thick]
\tikzstyle{new style 0}=[circle, fill=black, draw=black, scale=1.0, thick]
\tikzstyle{new style 2}=[circle, fill=gray, draw=black, scale=1.0, thick]
\tikzstyle{new style 3}=[fill=white, draw=black, shape=circle, minimum size=0.5mm, inner sep = 0pt]
\tikzstyle{new style 4}=[fill={rgb,255: red,39; green,39; blue,39}, draw=black, shape=circle, minimum size=2mm, inner sep = 0pt]
\tikzstyle{new style 5}=[fill={rgb,255: red,123; green,123; blue,123}, draw=black, shape=circle, minimum size=2mm, inner sep = 0pt]

\tikzstyle{whiteSmallStyle}=[circle, fill=white, draw=black, scale=0.2]
\tikzstyle{blackStyle}=[circle, fill=black, draw=black, scale=0.3]
\tikzstyle{greyStyle}=[circle, fill=gray, draw=black, scale=0.3]
\tikzstyle{invisibleStyle}=[inner sep=0, draw=none, fill=none, scale=0.01]

\tikzstyle{whiteSmallStyle1}=[circle, fill=white, draw=black, scale=0.35, thick]
\tikzstyle{blackStyle1}=[circle, fill=black, draw=black, scale=0.5, thick]
\tikzstyle{greyStyle1}=[circle, fill=gray, draw=black, scale=0.5, thick]
\tikzstyle{invisibleStyle1}=[inner sep=0, draw=none, fill=none, scale=0.005]

\tikzstyle{white}=[circle, fill=white, draw=black, scale=0.8]
\tikzstyle{black}=[circle, fill=black, draw=black, scale=1.2]
\tikzstyle{grey}=[circle, fill=gray, draw=black, scale=1.2]
\tikzstyle{none}=[inner sep=0, draw=none, fill=black, scale=0.1]

\tikzstyle{white1}=[circle, fill=white, draw=black, scale=0.5]
\tikzstyle{black1}=[circle, fill=black, draw=black, scale=0.85]
\tikzstyle{grey1}=[circle, fill=gray, draw=black, scale=0.85]
\tikzstyle{none}=[inner sep=0, draw=none, fill=black, scale=0.1]

\tikzstyle{red1}=[circle, draw=black, fill=red, scale=0.85]
\tikzstyle{blue1}=[circle, draw=black, fill=blue, scale=0.85]

\tikzstyle{white2}=[circle, fill=white, draw=black, scale=0.35]
\tikzstyle{black2}=[circle, fill=black, draw=black, scale=0.3]
\tikzstyle{grey2}=[circle, fill=gray, draw=black, scale=0.3]
\tikzstyle{none}=[inner sep=0, draw=none, fill=black, scale=0.01]

\tikzstyle{empty}=[circle, fill=none, draw=none, scale=0.55]

\tikzstyle{empty_large}=[circle, fill=none, draw=none, scale=1.55]

\tikzstyle{thin edge}=[-, draw=black, tikzit draw=none, tikzit fill={rgb,255: red,39; green,39; blue,39}, very thin]
\tikzstyle{new edge style 0}=[-, draw=black, tikzit draw=none, tikzit fill={rgb,255: red,39; green,39; blue,39}, thick]
\tikzstyle{new edge style 1}=[-, draw=black, tikzit draw=none, tikzit fill={rgb,255: red,39; green,39; blue,39}, very thick]
\tikzstyle{new edge style 2}=[-, draw=black, tikzit draw=none, tikzit fill={rgb,255: red,39; green,39; blue,39}, very thick]

\def\BibTeX{{\rm B\kern-.05em{\sc i\kern-.025em b}\kern-.08em
    T\kern-.1667em\lower.7ex\hbox{E}\kern-.125emX}}

\begin{document}

\title{Entanglement-Efficient Distribution of Quantum Circuits over Large-Scale Quantum Networks \\

}

\author{

\IEEEauthorblockN{Felix Burt\IEEEauthorrefmark{1}\IEEEauthorrefmark{2}, Kuan-Cheng Chen\IEEEauthorrefmark{1}\IEEEauthorrefmark{2}, Kin K. Leung\IEEEauthorrefmark{1}}
\IEEEauthorblockA{\IEEEauthorrefmark{1}Department of Electrical and Electronic Engineering, Imperial College London, London, UK}
\IEEEauthorblockA{\IEEEauthorrefmark{2}Centre for Quantum Engineering, Science and Technology (QuEST), Imperial College London, London, UK}
\IEEEauthorblockA{f.burt23@imperial.ac.uk}}

\maketitle

\thispagestyle{plain}
\pagestyle{plain}

\begin{abstract}

Quantum computers face inherent scaling challenges, a fact that necessitates investigation of distributed quantum computing systems, whereby scaling is achieved through interconnection of smaller quantum processing units. However, connecting large numbers of QPUs will eventually result in connectivity constraints at the network level, where the difficulty of entanglement sharing increases with network path lengths. This increases the complexity of the quantum circuit partitioning problem, since the cost of generating entanglement between end nodes varies with network topologies and existing links. We address this challenge using a simple modification to existing partitioning schemes designed for all-to-all connected networks, that efficiently accounts for both of these factors. We investigate the performance in terms of entanglement requirements and optimisation time of various quantum circuits over different network topologies, achieving lower entanglement costs in the majority of cases than state-of-the-art methods. We provide techniques for scaling to large-scale quantum networks employing both network and problem coarsening. We show that coarsened methods can achieve improved solution quality in most cases with significantly lower run-times than direct partitioning methods.

\end{abstract}


\section{Introduction}\label{sec:intro}

In recent years, a significant amount of attention has been placed on modular and distributed quantum computing architectures, in which large-scale quantum computers are built by connecting multiple, smaller quantum processing units (QPUs) using quantum links \cite{CALEFFI2024110672,barral_review_2024}. Quantum links allow entanglement to be shared between separated QPUs, such that qubits may be interacted over distance using teleportation procedures. As a result, qubit connectivity is constrained on two levels. At the \textit{intra-QPU} level, internal connectivity limits which qubits can directly interact. At the \textit{inter-QPU} level, qubits may be restricted to only interact with each other via entanglement-based teleportation protocols, on top of any internal qubit routing. Sharing entanglement is typically slower and noisier than SWAP-based internal routing \cite{isailovic_interconnection_2006,ARQUIN,main_distributed_2025}, indicating that compilers should target the inter-QPU level first when trying to minimise additional overhead. Various methods have been developed to this end, mostly concerned with homogeneous, or all-to-all connected quantum networks \cite{zomorodi-moghadam_optimizing_2018, daei_optimized_2020, dadkhah_new_2021,ferrari_compiler_2021,sundaram_efficient_2021,nikahd_automated_2021,wu_entanglement-efficient_2023,wu_autocomm_2022,wu_qucomm_2023,baker_time-sliced_2020,cuomo_optimized_2023,ferrari_modular_2023,crampton_genetic_2024,promponas_compiler_2024,chen_circuit_2024,sundaram2024, cha_module-conditioned_2025, kaur2025, russo_telesabre_2025, mengoni_efficient_2025}. This problem becomes more complicated when inter-QPU connectivity is constrained by quantum network topologies, causing the communication overhead to depend on the network path. The number of works that have tackled this problem is more sparse, and the solutions that currently exist \cite{sundaram_2022, andres-martinez_distributing_2024, sundaram2024, liu_ecdqc_2025} do not consider the full spectrum of possibilities for teleportation when partitioning and distributing quantum circuits. In this work, we extend the framework proposed in previous work (Ref. \cite{burt2025}) to general network topologies, using a novel technique to account for network dependent entanglement costs. Furthermore, we provide a means to scale to large-scale DQC systems using network coarsening techniques. Using network coarsening, solution quality is improved for linear networks, and competitive for grid networks compared with direct partitioning. Additionally, these results are achieved with significantly reduced run-times. 

\section{Background}\label{sec:background}

\subsection{Quantum teleportation}\label{sec:teleportation}

The backbone of distributed quantum computing is the ability to teleport qubits and gates across QPUs, achieved using a combination of shared entanglement and \textit{local operations and classical communication} (LOCC). The process of teleportation allows for the transfer of quantum information without the physical transmission of the qubit itself. Non-local two-qubit operations can be performed by teleporting qubits between QPUs, which we refer to as \textit{state teleportation}, or entangling qubits with auxiliary, communication qubits in distant QPUs, using the communication qubits to perform controlled-unitary operations. This is referred to as \textit{gate teleportation}. Both gate teleportation and state teleportation are achieved using the same primitive operations, the entanglement-assisted starting and ending processes \cite{eisert_optimal_2000, wu_entanglement-efficient_2023,yimsiriwattana_distributed_2004}. The starting process, denoted $S_{q,e}$, maps the state of a \textit{root} qubit $q$ onto an \textit{auxiliary} communication qubit $e$ in a distant QPU. For an input state $\ket{\psi}_{q} = \alpha \ket{0} + \beta \ket{1}$, the starting process performs the following map:

\begin{equation}\label{eq:starting}
    \begin{aligned}
        S_{q,e} (\ket{\psi}_{q})  &\to CX_{q,e} \ket{\psi}_{q}\ket{0}_{e} \\
        &= \alpha \ket{0}_{q}\ket{0}_{e} + \beta \ket{1}_{q}\ket{1}_{e},
    \end{aligned}
\end{equation}

where the effect of the $CX$ is achieved using the sub-circuit in Fig. \ref{fig:starting}. This process consumes a shared \textit{e-bit}, or \textit{EPR-pair}, which is a pair of entangled qubits assumed to be in a Bell state $\ket{\Phi^{+}} = \frac{1}{\sqrt{2}}(\ket{00} + \ket{11})$. The resulting state in Eq. \ref{eq:starting} allows $e$ to be used in place of $q$ for controlled-unitary operations, until the symmetry between $q$ and $e$ is broken by a non-diagonal single-qubit gate \cite{wu_entanglement-efficient_2023}. The ending process $E_{q,e}$ has the inverse effect of the starting process, and is used to disentangle the qubit $q$ from the communication qubit $e$, which is achieved using only LOCC. The ending process performs the following map:

\begin{equation}\label{eq:ending}
    \begin{aligned}
        & E_{q,e}(\ket{\psi'}_{q,e}) \\ &= Tr_{e}(CX_{q,e} \ket{\psi'}_{q,e} \bra{\psi'}_{q,e} CX_{q,e}),
    \end{aligned}
\end{equation}

% \begin{figure}
%     \centering
%     \resizebox{0.9\linewidth}{!}{
%     \tikzfig{Figures/start_end}
%     }
%     \caption{}
%     \label{fig:starting_ending}
% \end{figure}

\begin{figure}
    \centering
    \usetikzlibrary{decorations.pathmorphing}
\providecommand{\ket}[1]{\left|#1\right\rangle}
\begin{tikzpicture}[scale=1.000000,x=1pt,y=1pt]
\filldraw[color=white] (0.000000, -7.500000) rectangle (216.000000, 82.500000);
% Drawing wires
% Line 9: 0 W q_0
\draw[color=black] (20.000000,60.000000) -- (110.000000,60.000000);
%\draw[color=black] (0.000000,60.000000) -- (216.000000,60.000000);
%\draw[color=black] (0.000000,60.000000) node[left] {$q_1$};
\draw[color=black] (20.000000,60.000000) node[left] {$q$};
% Line 10: 1 W q_1
% \draw[color=black] (0.000000,60.000000) -- (216.000000,60.000000);
% \draw[color=black] (0.000000,60.000000) node[left] {$q_1$};
% Line 11: 4 W e_0
\draw[color=black] (32.000000,45.000000) -- (72.000000,45.000000);
\draw[color=black] (72.000000,44.500000) -- (96.000000,44.500000);
\draw[color=black] (72.000000,45.500000) -- (96.000000,45.500000);
% Line 12: 5 W e_1
\draw[color=black] (32.000000,20.000000) -- (110.000000,20.000000);
% \draw[color=black] (100.000000,19.500000) -- (124.000000,19.500000);
% \draw[color=black] (100.000000,20.500000) -- (124.000000,20.500000);
\draw[color=black] (20.000000,20.000000) node[left] {$e_0$};
% Line 13: 2 W q_2
% \draw[color=black] (0.000000,5.000000) -- (216.000000,5.000000);
% \draw[color=black] (0.000000,5.000000) node[left] {$q_2$};
% % Line 14: 3 W q_3
% \draw[color=black] (0.000000,-10.000000) -- (216.000000,-10.000000);
% \draw[color=black] (0.000000,-10.000000) node[left] {$q_3$};
% Done with wires; drawing gates
% Line 16: 0 G $H$
% \begin{scope}
% \draw[fill=white] (12.000000, 75.000000) +(-45.000000:8.485281pt and 8.485281pt) -- +(45.000000:8.485281pt and 8.485281pt) -- +(135.000000:8.485281pt and 8.485281pt) -- +(225.000000:8.485281pt and 8.485281pt) -- cycle;
% \clip (12.000000, 75.000000) +(-45.000000:8.485281pt and 8.485281pt) -- +(45.000000:8.485281pt and 8.485281pt) -- +(135.000000:8.485281pt and 8.485281pt) -- +(225.000000:8.485281pt and 8.485281pt) -- cycle;
% \draw (12.000000, 75.000000) node {$H$};
% \end{scope}
% Line 21: 4 epr 5
\draw[decorate,decoration={snake,amplitude=.4mm,segment length=2mm,post length=0.5mm}] (32.000000,45.000000) -- (32.000000,20.000000);
\filldraw (32.000000, 45.000000) circle(1.500000pt);
\filldraw (32.000000, 20.000000) circle(1.500000pt);
% Line 18: 0 +1
% \draw (33.000000,75.000000) -- (33.000000,60.000000);
% \filldraw (33.000000, 75.000000) circle(1.500000pt);
% \begin{scope}
% \draw[fill=white] (33.000000, 60.000000) circle(3.000000pt);
% \clip (33.000000, 60.000000) circle(3.000000pt);
% \draw (30.000000, 60.000000) -- (36.000000, 60.000000);
% \draw (33.000000, 57.000000) -- (33.000000, 63.000000);
% \end{scope}
% Line 22: 0 +4
\draw (51.000000,60.000000) -- (51.000000,45.000000);
\filldraw (51.000000, 60.000000) circle(1.500000pt);
\begin{scope}
\draw[fill=white] (51.000000, 45.000000) circle(3.000000pt);
\clip (51.000000, 45.000000) circle(3.000000pt);
\draw (48.000000, 45.000000) -- (54.000000, 45.000000);
\draw (51.000000, 42.000000) -- (51.000000, 48.000000);
\end{scope}
% % Line 23: 4 M
\draw[fill=white] (66.000000, 39.000000) rectangle (78.000000, 51.000000);
\draw[very thin] (72.000000, 45.600000) arc (90:150:6.000000pt);
\draw[very thin] (72.000000, 45.600000) arc (90:30:6.000000pt);
\draw[->,>=stealth] (72.000000, 39.600000) -- +(80:10.392305pt);
% Line 24: 5 X 4:owire
\draw (95.500000,45.000000) -- (95.500000,20.000000);
\draw (96.500000,45.000000) -- (96.500000,20.000000);
\begin{scope}
\draw[fill=white] (96.000000, 20.000000) +(-45.000000:8.485281pt and 8.485281pt) -- +(45.000000:8.485281pt and 8.485281pt) -- +(135.000000:8.485281pt and 8.485281pt) -- +(225.000000:8.485281pt and 8.485281pt) -- cycle;
\clip (96.000000, 20.000000) +(-45.000000:8.485281pt and 8.485281pt) -- +(45.000000:8.485281pt and 8.485281pt) -- +(135.000000:8.485281pt and 8.485281pt) -- +(225.000000:8.485281pt and 8.485281pt) -- cycle;
\draw (96.000000, 20.000000) node {$X$};
\end{scope}
\filldraw (96.000000, 45.000000) circle(1.500000pt);
\node at (130,35.5) {\large $=$};
% New wires after equals sign
\draw[color=black] (150.000000,60.000000) -- (200.000000,60.000000);
\draw[color=black] (175.000000,20.000000) -- (200.000000,20.000000);
\draw[->,decorate,decoration={snake,amplitude=.4mm,segment length=2mm,post length=1mm}] (175,60) -- (175,20);
\end{tikzpicture}
    \caption{The starting process $S_{q,e}$ that projects the state of a root qubit $q$ onto an auxiliary communication qubit $e$. This is the starting primitive for both state and gate teleportation.}
    \label{fig:starting}
\end{figure}
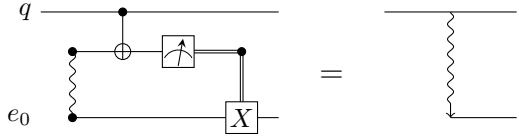

\begin{figure}
    \centering
    \usetikzlibrary{decorations.pathmorphing}
\providecommand{\ket}[1]{\left|#1\right\rangle}
\newcommand{\doublesnake}[3]{%
  \draw[decorate,decoration={snake,amplitude=.4mm,segment length=2mm,post length=1mm}] (#1,#2) -- (#1,#3);%
  \draw[decorate,decoration={snake,amplitude=.4mm,segment length=2mm,post length=1mm}] (#1+2,#2) -- (#1+2,#3);%
  \draw[->] (#1+1,#3+2) -- (#1+1,#3+2);%
}
\begin{tikzpicture}[scale=1.000000,x=1pt,y=1pt]
\filldraw[color=white] (0.000000, -7.500000) rectangle (216.000000, 82.500000);
% Drawing wires
% Line 9: 0 W q_0
\draw[color=black] (20.000000,60.000000) -- (110.000000,60.000000);
%\draw[color=black] (0.000000,60.000000) -- (216.000000,60.000000);
%\draw[color=black] (0.000000,60.000000) node[left] {$q_1$};
\draw[color=black] (20.000000,60.000000) node[left] {$q$};
% Line 10: 1 W q_1
% \draw[color=black] (0.000000,60.000000) -- (216.000000,60.000000);
% \draw[color=black] (0.000000,60.000000) node[left] {$q_1$};
% Line 11: 4 W e_0
% \draw[color=black] (32.000000,45.000000) -- (72.000000,45.000000);
% \draw[color=black] (72.000000,44.500000) -- (96.000000,44.500000);
% \draw[color=black] (72.000000,45.500000) -- (96.000000,45.500000);
% \draw[color=black] (20.000000,45.000000) node[left] {$e_0$};
% Line 12: 5 W e_1
\draw[color=black] (20.000000,20.000000) -- (70.000000,20.000000);
% \draw[color=black] (100.000000,19.500000) -- (124.000000,19.500000);
% \draw[color=black] (100.000000,20.500000) -- (124.000000,20.500000);
\draw[color=black] (20.000000,20.000000) node[left] {$e_0$};
\draw[fill=white] (66.000000, 39.000000-25) rectangle (78.000000, 51.000000-25);
\draw[very thin] (72.000000, 45.600000-25) arc (90:150:6.000000pt);
\draw[very thin] (72.000000, 45.600000-25) arc (90:30:6.000000pt);
\draw[->,>=stealth] (72.000000, 39.600000-25) -- +(80:10.392305pt);
% Line 24: 5 X 4:owire
\draw (96.500000-25,60.000000) -- (96.500000-25,26.000000);
\draw (97.500000-25,60.000000) -- (97.500000-25,26.000000);

\begin{scope}
\draw[fill=white] (97.000000-25, 60.000000) +(-45.000000:8.485281pt and 8.485281pt) -- +(45.000000:8.485281pt and 8.485281pt) -- +(135.000000:8.485281pt and 8.485281pt) -- +(225.000000:8.485281pt and 8.485281pt) -- cycle;
\clip (97.000000-25, 60.000000) +(-45.000000:8.485281pt and 8.485281pt) -- +(45.000000:8.485281pt and 8.485281pt) -- +(135.000000:8.485281pt and 8.485281pt) -- +(225.000000:8.485281pt and 8.485281pt) -- cycle;
\draw (97.000000-25, 60.000000) node {$Z$};
\end{scope}
% \filldraw (96.000000, 45.000000) circle(1.500000pt);
% % X gate on q_2

% % Line 28: 5 +2
% \draw (117.000000,20.000000) -- (117.000000,5.000000);
% \filldraw (117.000000, 20.000000) circle(1.500000pt);
% \begin{scope}
% \draw[fill=white] (117.000000, 5.000000) circle(3.000000pt);
% \clip (117.000000, 5.000000) circle(3.000000pt);
% \draw (114.000000, 5.000000) -- (120.000000, 5.000000);
% \draw (117.000000, 2.000000) -- (117.000000, 8.000000);
% \end{scope}
% Line 29: 5 +3
% \draw (135.000000,20.000000) -- (135.000000,-10.000000);
% \filldraw (135.000000, 20.000000) circle(1.500000pt);
% \begin{scope}
% \draw[fill=white] (135.000000, -10.000000) circle(3.000000pt);
% \clip (135.000000, -10.000000) circle(3.000000pt);
% \draw (132.000000, -10.000000) -- (138.000000, -10.000000);
% \draw (135.000000, -13.000000) -- (135.000000, -7.000000);
% \end{scope}
% % Line 31: 5 H
\begin{scope}
\draw[fill=white] (156.000000-105, 20.000000) +(-45.000000:8.485281pt and 8.485281pt) -- +(45.000000:8.485281pt and 8.485281pt) -- +(135.000000:8.485281pt and 8.485281pt) -- +(225.000000:8.485281pt and 8.485281pt) -- cycle;
\clip (156.000000-105, 20.000000) +(-45.000000:8.485281pt and 8.485281pt) -- +(45.000000:8.485281pt and 8.485281pt) -- +(135.000000:8.485281pt and 8.485281pt) -- +(225.000000:8.485281pt and 8.485281pt) -- cycle;
\draw (156.000000-105, 20.000000) node {$H$};
\end{scope}
% % Line 32: 5 M
% \draw[fill=white] (174.000000, 14.000000) rectangle (186.000000, 26.000000);
% \draw[very thin] (180.000000, 20.600000) arc (90:150:6.000000pt);
% \draw[very thin] (180.000000, 20.600000) arc (90:30:6.000000pt);
% \draw[->,>=stealth] (180.000000, 14.600000) -- +(80:10.392305pt);
% % Line 33: 1 Z 5:owire
% \draw (203.500000,60.000000) -- (203.500000,20.000000);
% \draw (204.500000,60.000000) -- (204.500000,20.000000);
% \begin{scope}
% \draw[fill=white] (204.000000, 60.000000) +(-45.000000:8.485281pt and 8.485281pt) -- +(45.000000:8.485281pt and 8.485281pt) -- +(135.000000:8.485281pt and 8.485281pt) -- +(225.000000:8.485281pt and 8.485281pt) -- cycle;
% \clip (204.000000, 60.000000) +(-45.000000:8.485281pt and 8.485281pt) -- +(45.000000:8.485281pt and 8.485281pt) -- +(135.000000:8.485281pt and 8.485281pt) -- +(225.000000:8.485281pt and 8.485281pt) -- cycle;
% \draw (204.000000, 60.000000) node {$Z$};
% \end{scope}
% \filldraw (204.000000, 20.000000) circle(1.500000pt);
% Done with gates; drawing ending labels
% Done with ending labels; drawing cut lines and comments
% Done with comments
% QPU labels at the start
% \node at (-15, 52.5) {$Q_0$};
% \node at (-15, 20) {$Q_1$};
% % Curly brackets for Q_0 and Q_1
% \draw[decorate,decoration={brace,amplitude=3pt}] (0,45) -- (0,60);
% \draw[decorate,decoration={brace,amplitude=3pt}] (0,15) -- (0,25);
% Dashed buffer line between e_0 and e_1
% \draw[dashed] (10.000000,32.500000) -- (120.000000,32.500000);

% \draw[dashed] (140.000000,32.500000) -- (180.000000,32.500000);
% Equals sign for circuit equivalence
\node at (130,35.5) {\large $=$};
% New wires after equals sign
\draw[color=black] (150.000000,60.000000) -- (200.000000,60.000000);
\draw[color=black] (150.000000,20.000000) -- (177.000000,20.000000);
% \doublesnake{175}{20}{58}
\draw[->,decorate,decoration={snake,amplitude=.4mm,segment length=2mm,post length=1mm}] (177.000000,20.000000) -- (177.000000,60.000000);

\end{tikzpicture}
    \caption{The ending process $E_{q,e}$ that disentangles the state of a qubit $q$ from an auxiliary communication qubit $e$, completing the teleportation process.}
    \label{fig:ending}

\end{figure}
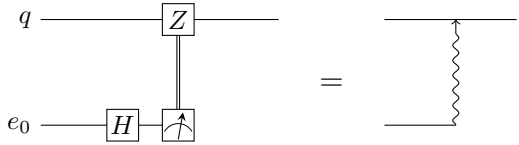

\begin{figure}
    \centering
    \usetikzlibrary{decorations.pathmorphing}
\providecommand{\ket}[1]{\left|#1\right\rangle}
\begin{tikzpicture}[scale=1.000000,x=1pt,y=1pt]
\filldraw[color=white] (0.000000, -7.500000) rectangle (96.000000, 142.500000);
% Drawing wires
% Line 10: x0 W q
\draw[color=black] (0.000000,135.000000) -- (96.000000,135.000000);
\draw[color=black] (0.000000,135.000000) node[left] {$q$};
% Line 11: y0 W e' color=white
\draw[color=white] (0.000000,120.000000) -- (9.000000,120.000000);
\draw[color=black] (9.000000,120.000000) -- (48.000000,120.000000);
\draw[color=black] (48.000000,119.500000) -- (87.000000,119.500000);
\draw[color=black] (48.000000,120.500000) -- (87.000000,120.500000);
% \draw[color=black] (0.000000,120.000000) node[left] {$e'$};
% Line 12: b0 W color=white
\draw[color=white] (0.000000,105.000000) -- (96.000000,105.000000);
% Line 13: z0 W e color=white
\draw[color=white] (0.000000,95.00000) -- (9.000000,95.00000);
\draw[color=black] (9.000000,95.00000) -- (101.00000,95.00000);
\draw[color=black] (0.000000,95.00000) node[left] {$e_{0}$};
% Ellipsis to indicate k repetitions
\draw (28.000000, 75.00000) node {$\vdots$};
% \draw (48.000000, 52.500000) node {$k$};
% Line 19: y2 W e' color=white
\draw[color=white] (0.000000,30.000000) -- (9.000000,30.000000+15);
\draw[color=black] (9.000000,30.000000++15) -- (48.000000,30.000000+15);
\draw[color=black] (48.000000,29.500000+15) -- (87.000000,29.500000+15);
\draw[color=black] (48.000000,30.500000+15) -- (87.000000,30.500000+15);
% \draw[color=white] (0.000000,30.000000) node[left] {$e'$};
% Line 20: b2 W color=white
\draw[color=white] (0.000000,15.000000) -- (96.000000,15.000000);
% Line 21: z2 W e color=white
\draw[color=white] (0.000000,0.000000+15) -- (9.000000,0.000000+15);
\draw[color=black] (9.000000,0.000000+15) -- (101.000000,0.000000+15);
\draw[color=black] (0.000000,0.000000+15) node[left] {$e_{k-1}$};
% Done with wires; drawing gates
% Line 24: y0 epr z0 y0:color=black z0:color=black
\draw[decorate,decoration={snake,amplitude=.4mm,segment length=2mm,post length=1mm}] (9.000000,120.000000) -- (9.000000,95.000000);
\filldraw (9.000000, 120.000000) circle(1.500000pt);
\filldraw (9.000000, 95.000000) circle(1.500000pt);
% Line 26: y2 epr z2 y2:color=black z2:color=black
\draw[decorate,decoration={snake,amplitude=.4mm,segment length=2mm,post length=1mm}] (9.000000,30.000000+15) -- (9.000000,0.000000+15);
\filldraw (9.000000, 30.000000+15) circle(1.500000pt);
\filldraw (9.000000, 0.000000+15) circle(1.500000pt);
% Line 28: +y0 x0
\draw (27.000000,135.000000) -- (27.000000,120.000000);
\begin{scope}
\draw[fill=white] (27.000000, 120.000000) circle(3.000000pt);
\clip (27.000000, 120.000000) circle(3.000000pt);
\draw (24.000000, 120.000000) -- (30.000000, 120.000000);
\draw (27.000000, 117.000000) -- (27.000000, 123.000000);
\end{scope}
\filldraw (27.000000, 135.000000) circle(1.500000pt);
% Line 30: +y2 z1
\draw (27.000000,45.000000+15) -- (27.000000,30.000000+15);
\begin{scope}
\draw[fill=white] (27.000000, 30.000000+15) circle(3.000000pt);
\clip (27.000000, 30.000000+15) circle(3.000000pt);
\draw (24.000000, 30.000000+15) -- (30.000000, 30.000000+15);
\draw (27.000000, 27.000000+15) -- (27.000000, 33.000000+15);
\end{scope}
% Line 32: y0 M
\draw[fill=white] (42.000000, 114.000000) rectangle (54.000000, 126.000000);
\draw[very thin] (48.000000, 120.600000) arc (90:150:6.000000pt);
\draw[very thin] (48.000000, 120.600000) arc (90:30:6.000000pt);
\draw[->,>=stealth] (48.000000, 114.600000) -- +(80:10.392305pt);
% Line 34: y2 M
\draw[fill=white] (42.000000, 24.000000+15) rectangle (54.000000, 36.000000+15);
\draw[very thin] (48.000000, 30.600000+15) arc (90:150:6.000000pt);
\draw[very thin] (48.000000, 30.600000+15) arc (90:30:6.000000pt);
\draw[->,>=stealth] (48.000000, 24.600000+15) -- +(80:10.392305pt);
% Line 39: y0 +y1 +y2 (segmented)
\draw (68.500000,120.000000) -- (68.500000,85.000000);
\draw (69.500000,120.000000) -- (69.500000,85.000000);
\filldraw (69.000000, 120.000000) circle(1.500000pt);
% Ellipsis in the middle
\draw (69.000000, 75.000000) node {$\vdots$};
% Continue from bottom
\draw (68.500000,45.000000+15) -- (68.500000,30.000000+15);
\draw (69.500000,45.000000+15) -- (69.500000,30.000000+15);
\begin{scope}
\draw[fill=white] (69.000000, 30.000000+15) circle(3.000000pt);
\clip (69.000000, 30.000000+15) circle(3.000000pt);
\draw (66.000000, 30.000000+15) -- (72.000000, 30.000000+15);
\draw (69.000000, 27.000000+15) -- (69.000000, 33.000000+15);
\end{scope}
% Line 41: +z0 y0:owire
\draw (86.500000,120.000000) -- (86.500000,95.000000);
\draw (87.500000,120.000000) -- (87.500000,95.000000);
\begin{scope}
\draw[fill=white] (87.000000, 95.000000) circle(3.000000pt);
\clip (87.000000, 95.000000) circle(3.000000pt);
\draw (84.000000, 95.000000) -- (95.000000, 95.000000);
\draw (87.000000, 87.000000) -- (87.000000, 98.000000);
\end{scope}
\filldraw (87.000000, 120.000000) circle(1.500000pt);
% Line 43: +z2 y2:owire
\draw (86.500000,30.000000+15) -- (86.500000,0.000000+15);
\draw (87.500000,30.000000+15) -- (87.500000,0.000000+15);
\begin{scope}
\draw[fill=white] (87.000000, 0.000000+15) circle(3.000000pt);
\clip (87.000000, 0.000000+15) circle(3.000000pt);
\draw (84.000000, 0.000000+15) -- (90.000000, 0.000000+15);
\draw (87.000000, -3.000000) -- (87.000000, 3.000000+15);
\end{scope}
\filldraw (87.000000, 30.00000+15) circle(1.500000pt);
% Done with gates; drawing ending labels
% Done with ending labels; drawing cut lines and comments
% Done with comments

\node at (130,75) {\large $=$};

\draw (150,135) -- (170,135);

\draw (150,95) -- (170,95);

\draw (160.000000, 75.000000) node {$\vdots$};

\draw (150.000000, 15.000000) -- (170.000000, 15.000000);

\draw[->,decorate,decoration={snake,amplitude=.4mm,segment length=2mm,post length=1mm}] (160,135) -- (160,95);

\draw[-,decorate,decoration={snake,amplitude=.4mm,segment length=2mm,post length=0mm}] (160,95) -- (160,85);

\draw[->,decorate,decoration={snake,amplitude=.4mm,segment length=2mm,post length=1mm}] (160,60) -- (160,15);

\end{tikzpicture}
    \caption{A $k$-fold starting process $S_{q,\mathbb{E}}$ on qubit $q$ and communication qubits $\mathbb{E} = \{e_{0}, e_{1}, \ldots, e_{k-1}\}$. The $i$-th correction operation $X_{e_{i}}$ is conditioned on the sum modulo $2$ of the measurements on qubits $\{e'_{0}, e'_{1}, \ldots, e'_{i}\}$.}
    \label{fig:k_fold_start}
\end{figure}
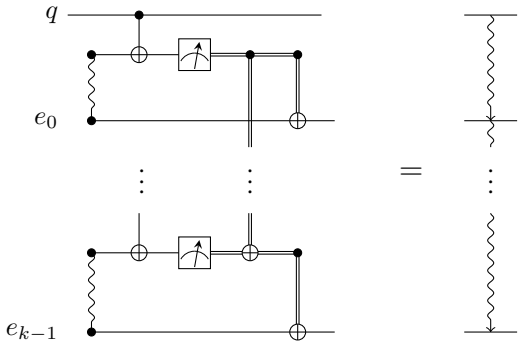

where $\ket{\psi'}_{q,e}$ is the joint state of $q$ and $e$. If $E_{q,e}$ is performed directly after $S_{q,e}$, the original state $\ket{\psi}_{q}$ is recovered. If a controlled-operation $CU_{e,q'}$ is performed on $e$ and a \textit{receiver} qubit $q'$ which is local to $e$, the ending process will map to the state:

\begin{equation}
    \begin{aligned}
        E_{q,e}(CU_{e,q'}\ket{\psi'}_{q,e} \ket{\phi}_{q'}) &= CU_{q',q} \ket{\psi}_{q} \ket{\phi}_{q'}, \\
    \end{aligned}
\end{equation}

such that the effect of a controlled-unitary from $q$ to $q'$ is achieved. Furthermore, if the ending process is performed from $q$, to $e$, we collapse the state onto $e$ instead of $q$, teleporting the state onto $e$:

\begin{equation}\label{eq:state_telep}
    E_{e,q} \circ S_{q,e}(\ket{\psi}_{q}) = \ket{\psi}_{e}.
  \end{equation}

Using this, a gate teleportation may be converted into a state teleportation, by changing the direction of the ending process after the controlled operation. In line with previous work, we refer to this as \textit{nested state teleportation} \cite{burt2025}.

A $k$-fold starting process $S_{q,\mathbb{E}}$, on a set of $k$ communication qubits $\mathbb{E} = \{e_{0},...,e_{k-1}\}$, links the qubit $q$ to all $k$ communication qubits. The direction of the final ending process determines the final location of the root qubit. Additionally, once the first starting process is performed, the communication qubit $e_{0}$ can be used as the root for the next starting process, meaning that any further starting process need not start again from $q$, and can make use of existing links. Furthermore, by delaying correction operations, the starting process can be performed on multiple communication qubits in parallel, allowing for a $k$-fold starting process to be performed in constant time. The correction $X$ operation for the $i$-th communication qubit $e_{i}$ uses the sum modulo $2$ of the measurements on qubits $\{e'_{0},e'_{1}..., e'_{i}\}$, as shown in Fig. \ref{fig:k_fold_start}. Note that, using this construction, all e-bits are requested between directly connected nodes, and no multi-hop communication in the network is required. Long-range entanglement is generated at the circuit level using the starting process. 

\subsection{Quantum circuit partitioning}\label{sec:qcp}

The high-level idea of quantum circuit partitioning is to split a quantum circuit into smaller sub-circuits that interact via shared entanglement and LOCC, in such a way that the entanglement required is minimised. The entanglement is expected to be delivered by the quantum network in the form of e-bits. The problem has been modelled in various ways, as a min-cut graph partitioning problem \cite{ferrari_modular_2023}, a hypergraph partitioning problem \cite{andres-martinez_automated_2019}, a vertex cover problem \cite{sundaram_efficient_2021, wu_entanglement-efficient_2023}, a quadratic assignment problem \cite{sundaram2024} a temporal partitioning problem \cite{baker_time-sliced_2020}, among others. In a limited number of cases, problem formulations have been extended to account for network topologies \cite{sundaram_distribution_2022,Andres-Martinez_2024,sundaram2024, liu_ecdqc_2025}, where the auxiliary e-bits required for multi-hop communications are considered. We focus on the temporal hypergraph formulation introduced in Refs. \cite{burt, burt2025}, that combines ideas from the hypergraph partitioning formulation of Andres-Martinez and Heunen \cite{andres-martinez_automated_2019} with the time-sliced partitioning from Baker et al. \cite{baker_time-sliced_2020} to achieve a general framework that considers multi-gate teleportation and state teleportation equally. In this formulation, a quantum circuit is transformed into a hypergraph $H(V,E)$, where the set of nodes $V$ corresponds to qubit time-step pairs $v = (q_{i}, t)$. Each node $v =(q_{i}, t)$ is connected to its temporal successor $v = (q_{i}, t+1)$ by an edge in $E$ except for nodes $v = (q_{i}, d-2)$, where $d$ is the depth, i.e., the final time-step, of the circuit. Since these edges connect the states of qubits at different time steps, they are referred to as \textit{state-edges}. Each node is associated with a gate, either a single-qubit gate (which may be an identity), or one qubit in a two-qubit gate. For convenience, we assumed the universal gate-set consisting of $U(\theta, \phi, \lambda)$ and $CP(\theta)$ gates. For each two-qubit gate between qubits $q_i$ and $q_j$, occurring at time $t$, we add an edge between nodes $(q_i, t)$ and $(q_j, t)$. 

Edges representing gates are then merged into hyper-edges, based on their \textit{compatibility} for gate teleportation. Gates are considered \textit{compatible} for gate teleportation if they have a common control qubit, and are only temporally separated by diagonal single-qubit gates or other $CP(\theta)$ gates on the common control qubit. This indicates that all gates in the group can be covered non-locally using $k$ e-bits if the target qubits are spread across $k$ QPUs excluding the QPU of the root qubit. Any target qubits local to the root qubit do not require an e-bit.

\begin{figure}
    \centering
    \input{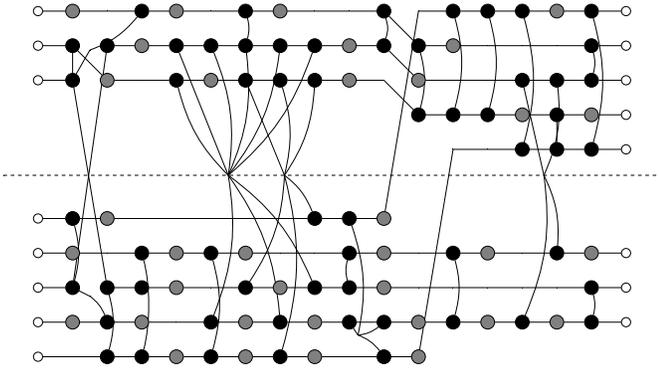}
    \caption{An example of a partitioned, temporal hypergraph. Gate teleportation is indicated by cut hyper-edges, while state teleportation is indicated by cut state-edges. Figure from \cite{burt2025}.}
    \label{fig:circuit_hypergraph}
\end{figure}

The cost of each hyper-edge is designed to correspond to the minimum number of e-bits required to cover all gates in the group. In order to define this, the nodes in each hyper-edge are split into two sets, $e_{root}$ and $e_{rec}$, where $e_{root}$ contains all nodes corresponding to the root-qubit over the time-span of the group. The nodes in $e_{rec}$ are the prospective ``receivers'', or target qubits in each group.

The hyper-edge cost is given by 

\begin{equation}\label{eq:cost}
    c_{e}(\Phi) = | \{\Phi(v) : v \in e_{rec} \} \setminus \{\Phi(u): u \in e_{root}\}|,
  \end{equation}

where $\Phi$ is the partition assignment function that assigns each node $v \in V$ to a QPU $Q_k \in Q$. Each $\Phi(v)$ from nodes in $e_{rec}$ indicates a QPU that requires a starting process. If $\Phi(v)$ is not in the set of QPUs assigned to $e_{root}$, then we require an e-bit to be generated between the QPUs. If $\Phi(v)$ is in the set of QPUs assigned to $e_{root}$, then a starting process is already accounted for by a cut state-edge between nodes in $e_{root}$. If $v_{max,e}$ denotes the root-node corresponding to the final time-step of the hyper-edge, then the final ending process must be routed to $\Phi(v_{max,e})$. All edges, including regular two-node edges are split into a root and receiver set, though for two-node edges the distinction is arbitrary, and the cost corresponds to the unweighted `cut' of the edge.

The overall objective of the problem is to choose a partition assignment function $\Phi$ that minimises the total cost of the hyper-edges,

\begin{equation}\label{eq:cost_full}
    \min_{\Phi} \sum_{e \in E} c_{e}(\Phi),
\end{equation}

while the data qubit capacity of each QPU is not exceeded for all $t$. The objective in Eq. \ref{eq:cost_full} corresponds to the number of end-to-end e-bits required between QPUs. This does not account for any varying connectivity between QPUs, essentially assuming all QPUs are directly connected. While this assumption may be reasonable for small-scale architectures, larger architectures are likely to have more complex topologies, where the auxiliary cost of generating e-bits between QPUs may vary. In this work, we extend the framework to account for these differences, and show how this can be efficiently integrated into partitioning heuristics.

\subsection{Multilevel partitioning with temporal coarsening}\label{sec:multi_level}

\begin{figure}
    \centering
    \resizebox{0.9\linewidth}{!}{
    \input{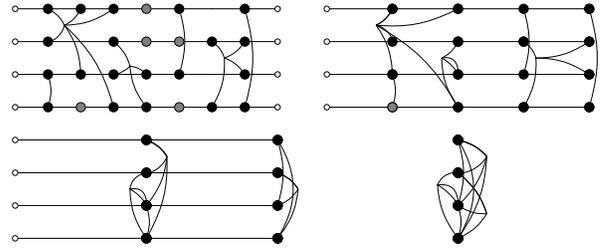}}
    \caption{An example of the recursive coarsening procedure for problem hypergraphs. The hypergraph is coarsened by merging pairs of temporally adjacent nodes, until the depth is reduced to $1$. The partitioning algorithm is then applied from the coarsest level to the finest level, refining the solution.}
    \label{fig:problem_coarsening}
\end{figure}

It is shown in Ref. \cite{burt2025} that the performance of partitioning algorithms for quantum circuits can be improved using a multilevel approach, in which problem hypergraphs are iteratively coarsened along the time axis and partitioned at each successive level. The most effective approach in previous work was a \textit{recursive} coarsening procedure that merges pairs of temporally adjacent nodes, as shown in Fig. \ref{fig:problem_coarsening}. This was shown to greatly improve solution quality as well as reduce run-time, using a tailored Fiduccia-Mattheyses (FM) heuristic \cite{fiduccia82} to refine the partitioning at each level. We refer the reader to Ref. \cite{burt2025} for a more detailed description of the coarsening procedure and the partitioning algorithm. 

\subsection{End-to-end entanglement distribution in quantum networks}\label{sec:end_to_end_entanglement}

The cost of generating entanglement between arbitrary end nodes in a quantum network depends on the network topology, since the shortest path between the two nodes may not be direct. We can model the network as a graph $G(Q,L)$, where $Q$ is our set of QPUs and the edge set is denoted by $L$, referring to direct links between QPUs. Each QPU is a set of physical data qubits $Q_{i} = \{\tilde{q}_{i,0}, \tilde{q}_{i,1}, ...\}$, where $i$ is the index of the QPU. Each edge $l_{ij}$ in $L$ connects QPUs $Q_i$ and $Q_j$. For simplicity, we consider the edge length to be uniform, such that the cost of generating an e-bit between directly connected nodes is equal. It was identified in Ref. \cite{andres-martinez_distributing_2024} that multi-QPU starting processes can be performed using a joint network path to reduce e-bit requirements, such that the auxiliary e-bit cost to connect $k$ QPUs corresponds to the \textit{minimum Steiner tree} connecting the QPUs in the network graph. Based on this, the authors propose a sub-routine for refining the output of a hypergraph partitioning algorithm by calculating Steiner trees across hyper-edges. In this work, we will extend this idea to the temporal hypergraph partitioning case, by generalising the Steiner tree problem to a \textit{Steiner forest} variant. While the Steiner tree and Steiner forest problems are NP-hard problems \cite{gassner_steiner_2010, biniaz_hardness_2015}, they can be trivial to solve for small instances. For larger cases, we will look to network coarsening techniques to simplify the problem.

\section{Temporal partitioning over general networks}\label{sec:temporal_partitioning_general}

\subsection{Generalising hyper-edge costs}\label{sec:hedge_costs}

In order to reflect the contribution of the network, it is necessary to generalise the edge costs in Eq. \ref{eq:cost}. We can do this using a generalisation of the Steiner tree problem, where we have a forest of trees corresponding to each QPU spanned by the root nodes. Let us first define the \textit{root and receiver partition sets} from edge $e$ under assignment $\Phi$, as

\begin{equation}
    \begin{aligned}
        \mathcal{P}_{e_{root},\Phi} &= \{\Phi(v) : v \in e_{root}\}, \\
        \mathcal{P}_{e_{rec},\Phi} &= \{\Phi(v) : v \in e_{rec}\}.
    \end{aligned}
\end{equation}

Each QPU $\in \mathcal{P}_{e_{root},\Phi}$, corresponds to a starting point for the entanglement generation, since it is eitherƒsub the initially assigned QPU of the root, or the receiver of a starting process accounted for by a cut state-edge. Our goal is to ensure that each QPU in $\mathcal{P}_{e_{rec},\Phi}$ is connected to at least one QPU in $\mathcal{P}_{e_{root},\Phi}$, via a tree. First, we calculate a Steiner tree connecting nodes in $\mathcal{P}_{e_{root},\Phi}$. This corresponds to the path along which the starting processes corresponding to state-edges will be performed (these are to be converted to nested state teleportations). Calling the set of nodes in this sub-graph $T_{e_{root}}$, we want to find a set of edges that connects each node in $\mathcal{P}_{e_{rec},\Phi}$ to at least one node in $T_{e_{root}}$. Calling the set of edges in the resulting forest $F_{e,\Phi}$, the cost of the hyper-edge is then given by the number of edges in the forest, $|F_{e,\Phi}|$ (that does not include the edges in $T_{e_{root}}$ since these are counted by the state edges), since this is the number of additional, auxiliary e-bits required. The equation for the cost is thus

\begin{equation}
    \begin{aligned}
        c_{e}(\Phi) &= |F_{e,\Phi}|, \\
    \end{aligned}
\end{equation}

making the overall objective

\begin{equation}\label{eq:cost_hetero}
    \min_{\Phi} \sum_{e \in E} |F_{e, \Phi}|.
\end{equation}

Note that this also counts the costs from the state-edges. Since state-edges are a special case of our hyper-edges with one node in $e_{root}$ and one in $e_{rec}$, the cost will simply be the path length from the root to the receiver. The state-edges connecting nodes in the root set of another hyper-edge will form a tree. Note that this tree will not necessarily be a Steiner tree for the nodes in the root set, since the path depends on $\Phi$. However, when optimising the assignment we will be favouring Steiner trees, since these will be the cheapest paths. Additionally, in most cases, it is unlikely that there will be more than two nodes in the root set of any hyper-edge, since the only possible advantage comes from the final ending process, which may or may not result in nested state teleportation. In this case, the tree will simply be the shortest path between the two nodes in $e_{root}$. 

\subsection{Adaptation of partitioning heuristics}\label{sec:adaptation}

We would like to be able to adapt existing partitioning heuristics to incorporate this cost without too much additional overhead. This is possible, provided the number of QPUs is not too large. Many partitioning heuristics rely on a \textit{gain structure} that stores local improvements to the overall cost that can be made by moving nodes to external partitions. For example, the Fiduccia-Mattheyses (FM) algorithm \cite{fiduccia82}, which is used in Ref. \cite{burt2025}, uses such a gain structure in order to choose moves, and efficiently maintains the gains by performing updates on neighbours of nodes that are moved. This can be complex if we need to recompute edges costs consistently, so we would like to store as many results as possible. Note that the costs are uniquely determined by the \textit{edge configurations}, i.e., which QPUs are present in the root and receiver partition sets. We can define a root and receiver configuration as a binary string of length $N$ for $N$ QPUs in the network. The root configuration of an edge $e$ is given by 

\begin{equation}
    cfg^{(e_{r})}_{i}(\Phi)= 
\begin{cases}
        1 & \text{if } \exists v \in e_{r} : \Phi(v) = i \\
        0 & \text{otherwise}
\end{cases}
\end{equation} 

where $r$ identifies either the root or receiver node set. Each root/rec configuration pair corresponds to a forest, and thus a cost. This means that there are $2^{2N}$ possible configurations for each hyper-edge, which is a large number. For moderate $N$, up to around $N=10$, we can pre-compute the costs of all edge configurations and store these in a lookup table \cite{huang_partitioning-based_1997}. Each gain update requires looking up no more than $4$ edge costs, $c_{e}(\Phi)$, $c_{e}(\Phi')$, $c_{e}(\tilde{\Phi})$ and $c_{e}(\tilde{\Phi})$. We use the $\Phi'$ to denote the updated assignment after node $v$ is moved, and $\tilde{\Phi}$ to denote the updated assignment after the neighbour, $u \in N(v)$, is moved, such that $\tilde{\Phi}'$ corresponds to the assignment after both $v$ and $u$ are moved. The contribution to the gain update from edge $e$ is given by 

\begin{equation}
    \begin{aligned}
        \Delta g_{u,v,e}(\Phi) &= c_{e}(\Phi') - c_{e}(\tilde{\Phi}') - c_{e}(\Phi) + c_{e}(\tilde{\Phi}),
    \end{aligned}
\end{equation}

such that the total gain update is the sum of the gains for all edges affected. This is given by

\begin{gather}\label{eq:delta_gain}
    \Delta g_{u,v}(\Phi) = \sum_{e \in A(u) \cap A(v)}[ c_{e}(\Phi') - c_{e}(\tilde{\Phi}') -  c_{e}(\Phi) + c_{e}(\tilde{\Phi})],
  \end{gather}

where $A(x)$ is the set of edges containing $x$. Provided we store the edge costs and configurations, and the number of edges per node is bounded, we can compute each gain update in constant time, such that the gain updates for all neighbours takes $\mathcal{O}(|N(v)|)$ time. This is essential for many partitioning heuristics, including  the tailored FM algorithm from Burt et al. \cite{burt2025}, which we will implement with this modification.

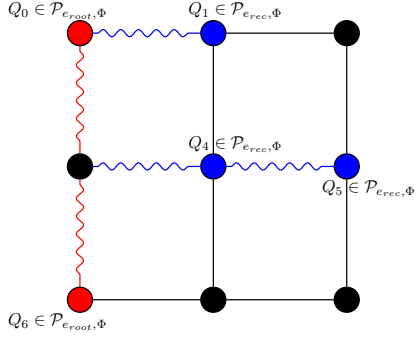
\begin{figure}
    \centering
    \resizebox{0.7\linewidth}{!}{
    \begin{tikzpicture}
	\begin{pgfonlayer}{nodelayer}
		\node [style=red1] (0) at (-7, 8) {};
		\node [style=black1] (1) at (-7, 5) {};
		\node [style=blue1] (2) at (-4, 8) {};
		\node [style=blue1] (3) at (-4, 5) {};
		\node [style=red1] (4) at (-7, 2) {};
		\node [style=black1] (5) at (-4, 2) {};
		\node [style=blue1] (6) at (-1, 5) {};
		\node [style=black1] (7) at (-1, 8) {};
		\node [style=black1] (8) at (-1, 2) {};
		\node [style=empty] (11) at (-7.5, 8.5) {$Q_{0} \in \mathcal{P}_{e_{root},\Phi}$};
		\node [style=empty] (14) at (-3.5, 5.5) {$Q_{4} \in \mathcal{P}_{e_{rec},\Phi}$};
		\node [style=empty] (15) at (-0.5, 4.5) {$Q_{5} \in \mathcal{P}_{e_{rec},\Phi}$};
		\node [style=empty] (16) at (-7.5, 1.5) {$Q_{6} \in \mathcal{P}_{e_{root},\Phi}$};
		\node [style=empty] (17) at (-3.5, 8.5) {$Q_{1} \in \mathcal{P}_{e_{rec},\Phi}$};
	\end{pgfonlayer}
	\begin{pgfonlayer}{edgelayer}
		\draw [decorate,decoration={snake,amplitude=.4mm,pre length = 0.5mm, segment length=2mm,post length=0.5mm}, color = red] (0) -- (1);
		\draw [decorate,decoration={snake,amplitude=.4mm,pre length = 0.5mm, segment length=2mm,post length=0.5mm}, color = red] (1) -- (4);
		\draw [decorate,decoration={snake,amplitude=.4mm,pre length = 0.5mm, segment length=2mm,post length=0.5mm}, color = blue] (1) -- (3);
		\draw [decorate,decoration={snake,amplitude=.4mm,pre length = 0.5mm, segment length=2mm,post length=0.5mm}, color = blue] (0) -- (2);
		\draw (3) to (2);
		\draw (2) to (7);
		\draw (7) to (6);
		\draw [decorate,decoration={snake,amplitude=.4mm,pre length = 0.5mm, segment length=2mm,post length=0.5mm}, color = blue](3) -- (6);
		\draw (6) to (8);
		\draw (3) to (5);
		\draw (5) to (8);
		\draw (5) to (4);
	\end{pgfonlayer}
\end{tikzpicture}
    }
    \caption{Forest in grid network. $Q_0$ and $Q_6$ are the root QPUs, which are already connected through state edges. The receiver QPUs $Q_1$, $Q_4$ and $Q_5$ are connected to the root QPUs by the forest. Only the edges in the forest, coloured in blue, are counted towards the cost of the hyper-edge.}
    \label{fig:grid_forest}
\end{figure}

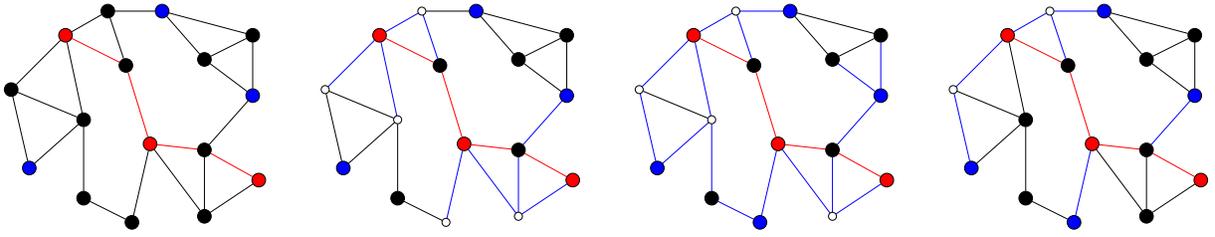
\begin{figure*}
    \centering
    \resizebox{0.9\linewidth}{!}{
    \begin{tikzpicture}
	\begin{pgfonlayer}{nodelayer}
		\node [style=black1] (0) at (4.75, 8) {};
		\node [style=red1] (1) at (5.75, 4.75) {};
		\node [style=black1] (2) at (8, 8.25) {};
		\node [style=black1] (3) at (8, 4.5) {};
		\node [style=black1] (4) at (3, 2.5) {};
		\node [style=white1] (5) at (3, 5.75) {};
		\node [style=white1] (6) at (8, 1.75) {};
		\node [style=red1] (7) at (10.25, 3.25) {};
		\node [style=white1] (8) at (5, 1.5) {};
		\node [style=blue1] (9) at (10, 6.75) {};
		\node [style=blue1] (10) at (6.25, 10.25) {};
		\node [style=white1] (11) at (4, 10.25) {};
		\node [style=white1] (12) at (0, 7) {};
		\node [style=red1] (13) at (2.25, 9.25) {};
		\node [style=blue1] (14) at (0.75, 3.75) {};
		\node [style=black1] (15) at (10, 9.25) {};
		\node [style=black1] (16) at (-8.25, 8) {};
		\node [style=red1] (17) at (-7.25, 4.75) {};
		\node [style=black1] (18) at (-5, 8.25) {};
		\node [style=black1] (19) at (-5, 4.5) {};
		\node [style=black1] (20) at (-10, 2.5) {};
		\node [style=black1] (21) at (-10, 5.75) {};
		\node [style=black1] (22) at (-5, 1.75) {};
		\node [style=red1] (23) at (-2.75, 3.25) {};
		\node [style=black1] (24) at (-8, 1.5) {};
		\node [style=blue1] (25) at (-3, 6.75) {};
		\node [style=blue1] (26) at (-6.75, 10.25) {};
		\node [style=black1] (27) at (-9, 10.25) {};
		\node [style=black1] (28) at (-13, 7) {};
		\node [style=red1] (29) at (-10.75, 9.25) {};
		\node [style=blue1] (30) at (-12.25, 3.75) {};
		\node [style=black1] (31) at (-3, 9.25) {};
		\node [style=black1] (32) at (17.75, 8) {};
		\node [style=red1] (33) at (18.75, 4.75) {};
		\node [style=black1] (34) at (21, 8.25) {};
		\node [style=black1] (35) at (21, 4.5) {};
		\node [style=black1] (36) at (16, 2.5) {};
		\node [style=white1] (37) at (16, 5.75) {};
		\node [style=white1] (38) at (21, 1.75) {};
		\node [style=red1] (39) at (23.25, 3.25) {};
		\node [style=blue1] (40) at (18, 1.5) {};
		\node [style=blue1] (41) at (23, 6.75) {};
		\node [style=blue1] (42) at (19.25, 10.25) {};
		\node [style=white1] (43) at (17, 10.25) {};
		\node [style=white1] (44) at (13, 7) {};
		\node [style=red1] (45) at (15.25, 9.25) {};
		\node [style=blue1] (46) at (13.75, 3.75) {};
		\node [style=black1] (47) at (23, 9.25) {};
		\node [style=black1] (48) at (30.75, 8) {};
		\node [style=red1] (49) at (31.75, 4.75) {};
		\node [style=black1] (50) at (34, 8.25) {};
		\node [style=black1] (51) at (34, 4.5) {};
		\node [style=black1] (52) at (29, 2.5) {};
		\node [style=black1] (53) at (29, 5.75) {};
		\node [style=black1] (54) at (34, 1.75) {};
		\node [style=red1] (55) at (36.25, 3.25) {};
		\node [style=blue1] (56) at (31, 1.5) {};
		\node [style=blue1] (57) at (36, 6.75) {};
		\node [style=blue1] (58) at (32.25, 10.25) {};
		\node [style=white1] (59) at (30, 10.25) {};
		\node [style=white1] (60) at (26, 7) {};
		\node [style=red1] (61) at (28.25, 9.25) {};
		\node [style=blue1] (62) at (26.75, 3.75) {};
		\node [style=black1] (63) at (36, 9.25) {};
	\end{pgfonlayer}
	\begin{pgfonlayer}{edgelayer}
		\draw [color = blue] (11) to (13);
		\draw [color = blue] (13) to (12);
		\draw (12) to (5);
		\draw [color = blue] (13) to (5);
		\draw (12) to (14);
		\draw (14) to (5);
		\draw [color = blue] (11) to (0);
		\draw [color = red] (0) to (13);
		\draw (11) to (10);
		\draw (10) to (2);
		\draw (2) to (15);
		\draw (15) to (10);
		\draw [color = red] (0) to (1);
		\draw (5) to (4);
		\draw (4) to (8);
		\draw [color = red] (1) to (3);
		\draw [color = blue] (6) to (7);
		\draw [color = red] (7) to (3);
		\draw [color = blue] (3) to (6);
		\draw [color = blue] (6) to (1);
		\draw [color = blue] (8) to (1);
		\draw (9) to (2);
		\draw (15) to (9);
		\draw [color = blue] (9) to (3);
		\draw (27) to (29);
		\draw (29) to (28);
		\draw (28) to (21);
		\draw (29) to (21);
		\draw (28) to (30);
		\draw (30) to (21);
		\draw (27) to (16);
		\draw [color = red] (16) to (29);
		\draw (27) to (26);
		\draw (26) to (18);
		\draw (18) to (31);
		\draw (31) to (26);
		\draw [color = red] (16) to (17);
		\draw (21) to (20);
		\draw (20) to (24);
		\draw [color = red] (17) to (19);
		\draw (22) to (23);
		\draw [color = red] (23) to (19);
		\draw (19) to (22);
		\draw (22) to (17);
		\draw (24) to (17);
		\draw (25) to (18);
		\draw (31) to (25);
		\draw (25) to (19);
		\draw [color = blue] (43) to (45);
		\draw [color = blue] (45) to (44);
		\draw (44) to (37);
		\draw [color = blue] (45) to (37);
		\draw [color = blue] (44) to (46);
		\draw [color = blue] (46) to (37);
		\draw [color = blue] (43) to (32);
		\draw [color = red] (32) to (45);
		\draw [color = blue] (43) to (42);
		\draw (42) to (34);
		\draw (34) to (47);
		\draw (47) to (42);
		\draw [color = red] (32) to (33);
		\draw [color = blue] (37) to (36);
		\draw [color = blue] (36) to (40);
		\draw [color = red] (33) to (35);
		\draw [color = blue] (38) to (39);
		\draw [color = red] (39) to (35);
		\draw [color = blue] (35) to (38);
		\draw [color = blue] (38) to (33);
		\draw [color = blue] (40) to (33);
		\draw [color = blue] (41) to (34);
		\draw [color = blue] (47) to (41);
		\draw [color = blue] (41) to (35);
		\draw [color = blue] (59) to (61);
		\draw [color = blue] (61) to (60);
		\draw (60) to (53);
		\draw (61) to (53);
		\draw [color = blue] (60) to (62);
		\draw (62) to (53);
		\draw [color = blue] (59) to (48);
		\draw [color = red] (48) to (61);
		\draw [color = blue] (59) to (58);
		\draw (58) to (50);
		\draw (50) to (63);
		\draw (63) to (58);
		\draw [color = red] (48) to (49);
		\draw (53) to (52);
		\draw (52) to (56);
		\draw [color = red] (49) to (51);
		\draw (54) to (55);
		\draw [color = red] (55) to (51);
		\draw (51) to (54);
		\draw (54) to (49);
		\draw [color = blue] (56) to (49);
		\draw (57) to (50);
		\draw (63) to (57);
		\draw [color = blue] (57) to (51);
	\end{pgfonlayer}
\end{tikzpicture}
    }
    \caption{Multi-sources BFS from nodes in the root tree. The initial root tree (red) forms the sources for the BFS. The receiver nodes (blue) are the terminals for the search. In the first iteration we explore the neighbours of all nodes in the root tree, which are accessible using one e-bit. Visited nodes are coloured white. Once all terminals are reached, we trace the shortest path back to the root tree and add remaining edges to the forest.}
    \label{fig:large_forest_BFS}
\end{figure*}

\subsection{Computing forests}\label{sec:forest_computation}

We can compute our forests using a multi-source breadth-first search (MSBFS), starting from each of our root nodes. At each iteration, we explore the neighbouring nodes from each of the root nodes, and add them to a queue. We then check if any of the nodes in the queue are in the receiver set. If so, we trace back the path to the root and add the edges to the forest. We continue until all nodes in the receiver set are connected to at least one node in the root set. The algorithm (Alg. \ref{alg:forest_bfs}) is as follows:

\begin{algorithm}
    \caption{Multi-source Breadth-First Search for Forest Computation}
    \label{alg:forest_bfs}
    \KwIn{Set of root nodes $R$, Set of receiver nodes $S$}
    \KwOut{Forest $F$ connecting $S$ to $R$}
    $F \gets \emptyset $\;
    Initialize queue $Q$\;
    \ForEach{$r \in R$}{
        Enqueue $Q$ with $r$\;
    }
    \While{$Q$ is not empty}{
        $current \gets Dequeue(Q)$\;
        \If{$current \in S$}{
            Trace back to root and add edges to $F$\;
        }
        \ForEach{neighbor $n$ of $current$}{
            Enqueue $Q$ with $n$\;
        }
    }
    \Return $F$\;
\end{algorithm}

If we are able to precompute all forests, then we add no additional pass complexity to the FM algorithm, and offload this all to the pre-computation. The limitation of doing this is that the pre-computation will become unfeasible for large quantum networks (reaching up to 20 QPUs), since the number of edge configurations scales exponentially. One obvious solution is to skip the precomputation, and simply build the look up table on the fly, storing edge costs as we compute them, but eventually this will lead to complex Steiner forest calculations that will be inefficient to compute within a gain update. An alternative option is to employ coarsening techniques at the network level in addition to the temporal coarsening of our problem graphs.

\section{Partitioning over large-scale quantum networks}\label{sec:large_scale}
In addition to the forest computation, direct $k$-way partitioning using FM has a factor $k$ in the complexity, which can make it slow for large networks. Many algorithms for $k$-way partitioning use a recursive approach, in which they first perform $l$-way partitioning for some $l<k$, then cut the graph into disconnected sub-graphs and partition the resulting sub-graphs. This can be repeated until we reach $k$ partitions, after which we can stitch together the results from each partition to form a complete solution. This avoids the $k$-scaling of direct partitioning and, and allows sub-graphs to be partitioned in parallel, leading to an exponential speedup in terms of $k$. For example, suppose we have a partitioning heuristic that scales as $\mathcal{O}(kn)$, where $k$ is the number of partitions and $n$ the number of nodes. At the first level, we take $l=2$ and partition into $2$ sub-graphs. This is done in $\mathcal{O}(n)$ time. Each sub-graph must be further partitioned into $k/2$ parts. At the next level, we partition $2$ graphs, each with roughly $n/2$ nodes. Since we can parallelise the partitioning, this also takes $\mathcal{O}(n)$ time. At the next level, we have $4$ graphs of $n/4$ nodes, and so on. After $log_{2}(k)$ levels, we have $k$ graphs of $n/k$ nodes. So the time taken is $\sum_{i=0}^{log_{2}(k)} \mathcal{O}(\frac{n}{2^{i}}) = \mathcal{O}(n)$. If we are unable to parallelise, we must partition sequentially at each level. Since the total number of nodes at each level always sums to $n$, the time taken is $\mathcal{O}(log_{2}(k)n)$. In both cases we get a strong speedup, though for some problem structures we may lose solution quality. 

\subsection{Network coarsening}\label{sec:net_coarse}
%%%Could shorten this
Since we are partitioning over a network, standard recursive partitioning would lead to losing key information about the network topology and, thus, the auxiliary entanglement costs. However, we can take inspiration from this idea to design an analogous approach. In Ref. \cite{burt2025}, the authors explored techniques for coarsening \textit{problem hypergraphs} corresponding to quantum circuits. It was shown that coarsening graphs along the time axis and partitioning using a multilevel approach led to faster runtimes and improved solution quality. This raises the question of whether similar techniques may be useful for simplifying networks over which we wish to partition our problem graphs. In addition to temporal coarsening of problem hypergraphs, we propose the use of \textit{network coarsening} routines. Coarsening a network, or any graph, typically involves identifying clusters or communities, or iteratively merging nodes together and contracting edges between them. By merging nearby nodes together, we can create coarser representations of the network, where each node contains a sub-network. We can then partition first at the level of the coarse network, and then partition different sections of the problem graph over different sub-networks.

To describe this process further, we first define the ``level'' of the network graph, $l$, to mean the coarsening stage. The original, fine-grained, network is thus $l=0$. Consider first a single level coarsening routine, starting with a network graph $G(Q, L)$ with $N = |Q|$ QPU nodes (note the difference between this quantity and $|Q_{i}|$, which is the number of qubits in QPU). We can coarsen the network by merging nodes together, such that we have $N_{max}$ nodes in the coarsened network. For each pair of merged nodes $Q_{i}$ and $Q_{j}$, we create a super node with qubits $Q_{i} \cup Q_{j}$ and drop the edge between them. To decide which merges to perform, we can use a matching algorithm that finds a set of edges in the graph such that no two edges share a node. This is done by iteratively computing a maximum weight matching of the graph and contracting edges in the matching until we reach the desired size. In order to encourage the merging of similar sizes nodes, we assign the following weight to each edge:

\begin{equation}
    w_{ij} = -(|Q_{i}| - |Q_{j}|)^{2},
\end{equation}

such that the weight of edges is maximised when the two nodes are of similar size. Nodes of different sizes will have a negative weight. We compute the matching using the built in networkX function \texttt{max\_weight\_matching} \cite{networkX}, which uses the blossom algorithm from Edmonds \cite{Edmonds_1965}. This runs in time $\mathcal{O}(n^{3})$, where $n$ is the number of nodes in the graph. We describe this in Alg. \ref{alg:coarsen_network}.

\begin{algorithm}[ht]
    \DontPrintSemicolon
    \SetAlgoLined
    \KwIn{A network graph $G$, and a desired size $N_{max}$}
    \KwOut{A coarsened network $G'$ with $N_{max}$ nodes}
    \caption{\textsc{CoarsenNetwork}($G, N_{max}$)}
    \label{alg:coarsen_network}
    \Begin{
        $G' \leftarrow G$\;
    
        $N \leftarrow \text{nodeCount}(G')$\$;
    
        \While{$N > N_{max}$}{
            $M \leftarrow \textsc{ComputeMatching}(G')$\;
            \If{$M = \emptyset$}{
                \textbf{break} \; \tcp*[r]{No more valid merges}
            }
    
            \ForEach{ Edge $e=(u,v)$ in $M$ }{
                $\textsc{MergeNodes}(G', u, v)$\;
                \If{$N > N_{max}$}{
                    \textbf{return} $(G')$ \; \tcp*[r]{Reached desired size}
                }
            }
            $N \leftarrow \text{number of nodes in }G'$\;
        }
    
        \textbf{return} $(G')$\;
    }
    \end{algorithm}

\begin{figure*}[ht]
    \centering
    \resizebox{0.8\linewidth}{!}{
    \input{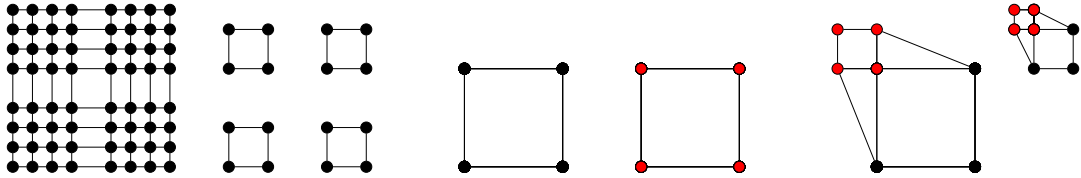}
    }
    \caption{Recursive coarsening of a grid network. The original network contains $N=64$ QPUs, and a coarsening factor $\chi=4$ is used. A single branch of the uncoarsening phase is shown on the right, where the active nodes are coloured red.}
    \label{fig:grid_coarse}
    \end{figure*}

\subsection{Sub-graph decomposition}

The network coarsening routine allows us to partition our problem hypergraph first over a simpler, coarsened network. Note that this network coarsening is completely independent of the coarsening of our problem hypergraph. This is because we only coarsen our problem hypergraphs temporally, such that we effectively reduce the depth of the circuit we are dealing with. The network coarsening affects the number of qubits in each QPU, and the paths between them, but has no effect on the problem hypergraphs. We can thus use a combination of temporal coarsening and network coarsening to simplify the problem hypergraphs and the networks, respectively. 
After partitioning over a coarsened network, we must cut the problem graphs into independent sub-graphs. With no network constraints, we could do this by simply dropping all cut edges. However, when we coarsen the network, we are reducing the paths lengths between nodes, and thus the cost of non-local edges. As such, the cost of a non-local edge may still change at lower level partitioning, since the nodes could be moved further away from each other on the overall network graph when partitioned independently at later levels. In order to account for this, we introduce \textit{dummy nodes}representing the QPUs outside the internal sub-network. Instead of dropping the non-local edges, we merge external nodes together into dummy nodes representing external partitions. Edges to dummy nodes represent connections to each other sub-graph, and their cost should be calculated in the same way as others. The use of dummy nodes means that all moves are aware of the global contribution to the cost, even if they only account for a part of the network path between two nodes.
Starting with a network of $N$ QPUs, consider coarsening the network down to $k$ QPUs. We then $k$-partition the graph over the coarse network. We then make $k$ copies of the coarse network, and for each copy, we partially uncoarsen one of the networks into $k$ QPUs, keeping the connections to the other, coarse nodes. Each sub-network now has $2k-1$ nodes, $k$ from the previous level, and $k-1$ which have just been uncoarsened. We then make $k$ copies of the problem hypergraph. Using our partitioning from the coarse level, each node in the problem hypergraph node set $V$ is assigned to some $Q_{i}$ according to the optimised assignment $\Phi$. For each of our $k$ copies, we keep all the nodes that are assigned to a particular $Q_{i}$, and, for all nodes outside, we merge them into a dummy node. This means that each problem graph will now have roughly $n/k + k$ nodes. We then partition each of these sub-graphs over the $2k-1$ QPUs in the corresponding sub-network keeping the dummy nodes locked in place. This way, dummy nodes do not contribute to the run-time of the partitioning, but still contribute to the global cost of each hyper-edge. This encourages nodes to not stray too far from their neighbours in other sub-graphs, while not directly calculating the cost over the full network. After partitioning each sub-graph, we can reconstruct a solution for the level $0$ graph by stitching together each of the optimised assignments from the sub-graphs.
\begin{algorithm}[ht]
    \DontPrintSemicolon
    \SetAlgoLined
    \KwIn{A network $G$, a parent network $\tilde{G}$, and a node to expand $v$}
    \KwOut{A sub-network $G'$ with node $v$ expanded}
    \caption{\textsc{ExpandNode}($G, \tilde{G}, v$)}
    \Begin{
        $G' \leftarrow G$\;

        $\tilde{V} \gets \textsc{NodesContainedIn}(v)$\; 
        % \tcp*[r]{Find nodes from parent network $\tilde{G}$ which are contained in $v$.}
        \For{each $u \in \tilde{V}$}{
            $G' \leftarrow \textsc{AddNode}(G', u)$\;
            \For{each $w \in N_{\tilde{G}}(u)$}{
                $w' \gets \textsc{ParentNode}(\tilde{G},G,w)$\;
                $G' \leftarrow \textsc{AddEdge}(G', u, w')$ \;
            } 
            % \tcp*[r]{Find the corresponding node to $w$ in $G$ and add edge to $G'$.}
        }
        \tcp*[r]{Merge all nodes from two levels prior}

    }
    \end{algorithm}

\begin{algorithm}[ht]
    \DontPrintSemicolon
    \SetAlgoLined
    \KwIn{A problem hypergraph $H(V, E)$, a network $G(Q, L)$, a partially uncoarsened sub-network $G'(Q', L')$, and an assignment function $\Phi$}
    \KwOut{$|G|$ smaller hypergraphs $H_{i}'$ corresponding to sub-network $G'$.}
    \caption{\textsc{CutHyperGraph}($H, G, G', \Phi$)}
    \label{alg:cut_hypergraph}
    \Begin{
        $HList \gets []$\;
        \For{each $Q_{i} \in Q$}{
            $H' \gets \textsc{Copy}(H)$\;
            \For{each $Q_{j} \in Q$ such that $Q_{j} \neq Q_{i}$}{
                $\textsc{AddDummyNode}(H', Q_{j})$\;
            }
            \For{each $v \in V$ such that $\Phi(v) \neq Q_{i}$}{
                    $H' \gets \textsc{MergeIntoDummy}(H', \Phi(v))$\;
                }
            $\text{Append } H' \text{ to } HList$\;
            }

        \textbf{return} $HList$\;

    }
    \end{algorithm}

\subsection{Recursive network coarsening}

We can place these sub-routines within a larger, recursive partitioning routine, where we have multiple levels of coarsening and cutting. Since the network coarsening is independent of the problem hypergraph coarsening, we can perform temporal coarsening and multilevel partitioning for each sub-graph, merging nodes in and out of dummies where necessary. Starting with a network of $G(Q, L)$ of $N$ QPUs and a problem hypergraph $H(V, E)$ of $n$ nodes, we define a coarsening factor $\chi$, which determines the size reduction at each level. We use this to determine the desired size for the next level via $k = \lfloor N / \chi \rfloor$ QPUs. We repeat the coarsening recursively until the number of nodes in the network is less than or equal to $\chi$. This process is described in Alg. \ref{alg:coarsen_network_recursive}.
Once we have all network levels, we proceed to $k$-partition the problem hypergraph at the coarsest level. We then make $k$ copies of the network at the current level and expand one of the sub-networks for each copy, marking the other $k-1$ sub-networks as dummy QPUs. If we have dummy nodes from the previous level, we merge them into one of the other $k-1$ sub-networks where possible. If there is no direct edge connecting the dummy QPUs from the previous level to the current level, we keep an additional dummy node from the previous level, such that we may, at worst, accumulate an extra $QPU$ per level. This means that the maximum number of QPUs in the sub-network is $2k - 1 + l$. We then cut the problem hypergraph into $k$ sub-graphs according to Alg. \ref{alg:cut_hypergraph}, resulting in $k$ graphs of roughly $n/k$ nodes. We repeat this process of partitioning, cutting problem graphs and partially uncoarsening the network until we reach the finest level at approximately level $l_{max} = log_{\chi}(N)$.

\begin{algorithm}[ht]
    \DontPrintSemicolon
    \SetAlgoLined
    \KwIn{An initial network $\textit{G(Q, L)}$ with node set $Q$ and edge set $L$, and a problem hypergraph $H(V, E)$ with node set $V$ and edge set $E$, a factor $l$}
    \KwOut{A list of successively coarsened networks $\textit{GList}$ and hypergraphs $\textit{HList}$}
    \caption{\textsc{CoarsenNetworkRecursive}($G, H, l$)}
    \label{alg:coarsen_network_recursive}
    \Begin{
        $G' \leftarrow G$\;
        $H' \leftarrow H$\;
        $k \leftarrow \text{nodeCount}(G')$\;
        $N_{max} \leftarrow \lfloor k / l \rfloor$\;
    
        % \tcp{Each node initially contains only itself}
        % $containedNodes \leftarrow \{\ n \mapsto \{n\} \mid n \in \text{nodes}(G')\}$\;
    
        $\textit{GList} \leftarrow [\,G\,]$\;
        $\textit{HList} \leftarrow [\,H\,]$\;
    
        \While{$k > l$}{
            $G' \leftarrow \textsc{CoarsenNetwork}(G',\, N_{max})$\;
            $H' \leftarrow \textsc{CoarsenHyperGraph}(H',\, N_{max})$\;
    
            $\textit{GList}.\text{append}(G')$\;
            $\textit{HList}.\text{append}(H')$\;
            % $networkCoarse.\textit{mapping} \leftarrow mapping$\;
    
            $k \leftarrow N_{max}$\;
            $N_{max} \leftarrow \lfloor k / l \rfloor$\;
            % $currentMapping \leftarrow \text{copy}(mapping)$\;
        }
    
        \textbf{return} $\textit{GList}, \textit{HList}$\;
    }
    \end{algorithm}

% \begin{figure}
%     \centering
%     \resizebox{0.9\linewidth}{!}{
%     \tikzfig{Figures/linear_coarse}
%     }
%     \caption{Recursive coarsening of a linear network.}
%     \label{fig:linear_coarse}
% \end{figure}

\section{Results} \label{sec:results}

We evaluate the performance of the heterogeneous partitioning scheme on a variety of circuits from the QASM benchmark suite \cite{li2022qasmbenchlowlevelqasmbenchmark} and fixed-depth random circuits with a varying two-qubit gate proportion, called $CP$-fraction \cite{burt,sundaram_efficient_2021} . This evaluation is split into two main parts. First, we use the set of benchmark circuits to evaluate the performance of our algorithm on each network topology, where network sizes are constrained to be have $12$ or fewer QPUs. We refer to these as \textit{intermediate-scale} quantum networks. For these cases, we use direct, $k$-partitioning, with recursive temporal coarsening but no network coarsening. We compare the performance with a number of methods for heterogeneous partitioning from the \textit{Pytket DQC} library \cite{andres-martinez_cqclpytket-dqc_2024,andres-martinez_distributing_2024}. We then evaluate the performance of recursive network coarsening for networks reaching up to 64 QPUs. We compare the best achievable with results with and without network coarsening.

\subsection{Network topologies}

We use a variety of simple network topologies to begin with, namely \textit{linear} and \textit{grid} networks. In addition, we consider random networks, produced using the \textit{Erdos-Renyi} model. 

\subsubsection{Linear}

Linear networks consist of $N$ QPUs in a line, with each QPU connected to its nearest neighbours.

\subsubsection{Grid}

A grid network consists of $N$ QPUs, arranged in a $2D$ grid. Each node is connected to its nearest neighbours.

\subsubsection{Random}

Random networks are generated using the \textit{Erdos-Renyi} model, where each node is connected to each other with a probability $p$. We post-select randomly generated networks on the condition that there are no disconnected nodes.

\subsection{Intermediate-scale networks}\label{sec:intermediate-scale-res}

\begin{figure*}[!htbp]
    \centering
    \begin{subfigure}[b]{0.32\textwidth}
        \centering
        \resizebox{\textwidth}{!}{
        \begin{tikzpicture}
\begin{axis}[
    width=10cm,
    height=8cm,
    % width = 0.2\textwidth,
    view={-45}{40},           % Adjust to change your 3D perspective
    xlabel={Fraction},
    ylabel={Number of qubits (Number of QPUs)},
    zlabel={Entanglement cost},
    grid=major,
    domain=0:1,
    colormap/viridis,
    xtick= {0.1,0.3,0.5,0.7,0.9},
    ytick={32,48,64,80,96},
    ztick={0,1000,2000,3000},
    yticklabels = {32 (4),48 (6),64 (8),80 (10),96 (12)},
    x dir=reverse,
  % y dir=reverse,
  % z dir=reverse,
]
  % The file "all_fractions.dat" should contain columns:
  % fraction  num_qubits  r_mean  r_min  r_max
  \addplot3 [
    surf,
    opacity=0.3,
    shader=flat,
    color= blue,
    mesh/cols=9,
    mesh/ordering=colwise,  % Tells pgfplots how to “connect” the data
  ] 
  table[
    x=fraction,
    y=num_qubits,
    z=r_cost_mean,
    col sep=space,
    header=has colnames,
  ]{PGF/CP/Linear/Linear_CP_2-12.dat};

  \addlegendentry{Temporal Partitioning FM}

  \addplot3 [
    surf,
    opacity=0.3,
    color=yellow,
    % shader=flat,
    mesh/cols=9,
    mesh/ordering=colwise,  % Tells pgfplots how to “connect” the data
    colormap/hot,
  ] 
  table[
    x=fraction,
    y=num_qubits,
    z=PE_cost_mean,
    col sep=space,
    header=has colnames
  ]{PGF/CP/Linear/Linear_CP_2-12_tket.dat};

  \addlegendentry{Pytket PartitionEmbed}

\end{axis}
\end{tikzpicture}}
        \caption{Linear network}
        \label{fig:linear_CP}
    \end{subfigure}
    \hfill
    \begin{subfigure}[b]{0.32\textwidth}
        \centering
        \resizebox{\textwidth}{!}{
        \begin{tikzpicture}
\begin{axis}[
    width=10cm,
    height=8cm,
    % width = 0.3\textwidth,
    view={-45}{40},           % Adjust to change your 3D perspective
    xlabel={Fraction},
    ylabel={Number of qubits (Number of QPUs)},
    zlabel={Entanglement cost},
    grid=major,
    domain=0:1,
    colormap/viridis,
    xtick= {0.1,0.3,0.5,0.7,0.9},
    ytick={32,48,64,80,96},
    ztick={0,1000,2000},
    yticklabels = {32 (4),48 (6),64 (8),80 (10),96 (12)},
    x dir=reverse,
  % y dir=reverse,
  % z dir=reverse,
]
  % The file "all_fractions.dat" should contain columns:
  % fraction  num_qubits  r_mean  r_min  r_max
  \addplot3 [
    surf,
    opacity=0.3,
    shader=flat,
    color= blue,
    mesh/cols=9,
    mesh/ordering=colwise,  % Tells pgfplots how to “connect” the data
  ] 
  table[
    x=fraction,
    y=num_qubits,
    z=r_cost_mean,
    col sep=space,
    header=has colnames,
  ]{PGF/CP/Grid/Grid_CP_2-12.dat};

  \addlegendentry{Temporal Partitioning FM}

  \addplot3 [
    surf,
    opacity=0.3,
    color=yellow,
    % shader=flat,
    mesh/cols=9,
    mesh/ordering=colwise,  % Tells pgfplots how to “connect” the data
    colormap/hot,
  ] 
  table[
    x=fraction,
    y=num_qubits,
    z=PE_cost_mean,
    col sep=space,
    header=has colnames
  ]{PGF/CP/Grid/Grid_CP_2-12_tket.dat};

  \addlegendentry{Pytket PartitionEmbed}

\end{axis}
\end{tikzpicture}}
        \caption{Grid network}
        \label{fig:grid_CP}
    \end{subfigure}
    \hfill
    \begin{subfigure}[b]{0.32\textwidth}
        \centering
        \resizebox{\textwidth}{!}{
        \begin{tikzpicture}
\begin{axis}[
    width=10cm,
    height=8cm,
    % width = 0.2\textwidth,
    view={-45}{40},           % Adjust to change your 3D perspective
    xlabel={Fraction},
    ylabel={Number of qubits (Number of QPUs)},
    zlabel={Entanglement cost},
    grid=major,
    domain=0:1,
    colormap/viridis,
    xtick= {0.1,0.3,0.5,0.7,0.9},
    ytick={32,48,64,80,96},
    ztick={0,1000,2000},
    yticklabels = {32 (4),48 (6),64 (8),80 (10),96 (12)},
    x dir=reverse,
  % y dir=reverse,
  % z dir=reverse,
]
  % The file "all_fractions.dat" should contain columns:
  % fraction  num_qubits  r_mean  r_min  r_max
  \addplot3 [
    surf,
    opacity=0.3,
    shader=flat,
    color= blue,
    mesh/cols=9,
    mesh/ordering=colwise,  % Tells pgfplots how to “connect” the data
  ] 
  table[
    x=fraction,
    y=num_qubits,
    z=r_cost_mean,
    col sep=space,
    header=has colnames,
  ]{PGF/CP/Random/Random_CP_2-12.dat};

  \addlegendentry{Temporal Partitioning FM}

  \addplot3 [
    surf,
    opacity=0.3,
    color=yellow,
    % shader=flat,
    mesh/cols=9,
    mesh/ordering=colwise,  % Tells pgfplots how to “connect” the data
    colormap/hot,
  ] 
  table[
    x=fraction,
    y=num_qubits,
    z=PE_cost_mean,
    col sep=space,
    header=has colnames
  ]{PGF/CP/Random/Random_CP_2-12_tket.dat};
  \addlegendentry{Pytket PartitionEmbed}

\end{axis}
\end{tikzpicture}}
        \caption{Random network}
        \label{fig:random_CP}
    \end{subfigure}
    \caption{Entanglement costs for varying circuit size and two-qubit gate fraction across different network topologies. The blue surface is below the yellow until the intersection around $0.6$, where the Pytket-DQC methods outperform the temporal partitioning scheme.}
    \label{fig:combined_CP}
\end{figure*}
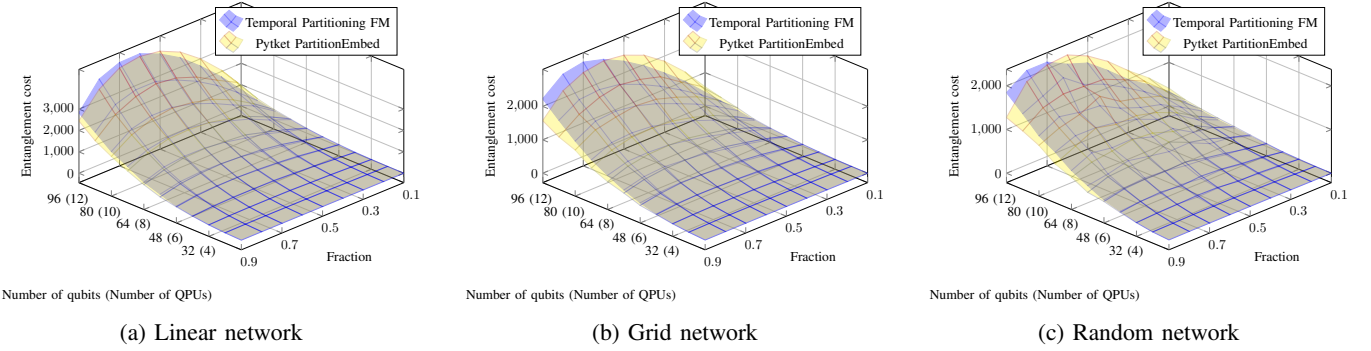

\begin{table*}[ht]
\centering
\begin{minipage}{0.48\textwidth}
\centering
\centering
\caption{QASM results for direct partitioning on linear networks.}
\label{tab:QASM_res_linear}
\resizebox{\linewidth}{!}{%
    \begin{filecontents*}{PGF/QASM/merged_results_linear_nq_50.dat}
circuit_name	num_qubits	num_partitions	pe_cost	pe_time	esd_cost	esd_time	r_cost	r_time
QV100	100	6	0	0	0	0	10984.8	133.34477696418762
QV100	100	8	0	0	0	0	16010.000	174.598
adder64	64	6	34.0	1.1453094005584716	30.6	6.643855476379395	21.4	56.91470818519592
adder64	64	8	48.4	1.1755083084106446	44.0	6.679545450210571	20.6	84.8583263874054
bv70	70	6	4.8	0.09510555267333984	5.0	0.172430419921875	5.0	9.258367395401
bv70	70	8	5.8	0.15089588165283202	5.8	0.2081925392150879	6.0	13.010530853271485
cat65	65	6	34.2	0.09516444206237792	34.4	0.17481722831726074	5.0	12.557146978378295
cat65	65	8	46.2	0.08972206115722656	59.6	0.18861861228942872	7.0	17.96216673851013
cc64	64	6	5.0	0.21776375770568848	5.0	0.3480709075927734	5.0	34.1944703578949
cc64	64	8	7.0	0.23661575317382813	7.0	0.3318434715270996	7.0	48.256580638885495
dnn51	51	6	102.2	1.4489152431488037	102.2	5.052017593383789	31.2	26.867007398605345
dnn51	51	8	121.2	2.03445782661438	122.6	7.67330846786499	32.8	35.81900172233581
ghz78	78	6	9.0	0.06958904266357421	9.0	0.13946294784545898	5.0	19.302638912200926
ghz78	78	8	34.6	0.1547306537628174	33.4	0.24226179122924804	7.0	28.145309162139892
ising66	66	6	6.0	0.13544554710388185	6.0	0.5590143203735352	5.0	0.8668134689331055
ising66	66	8	46.0	0.17243943214416504	46.4	0.5932968616485595	7.0	1.2750811100006103
ising98	98	6	28.6	0.3199711799621582	25.0	1.5362723827362061	5.0	1.3412904739379883
ising98	98	8	51.4	0.32952003479003905	48.8	1.524510955810547	7.0	1.977628469467163
knn67	67	6	14.8	0.8432032108306885	17.0	6.337424230575562	5.0	56.37756681442261
knn67	67	8	19.8	1.7488837718963623	23.2	6.610300874710083	8.2	88.09366383552552
qft63	63	6	96.0	17.51584358215332	96.0	88.17577605247497	150.0	132.098450422287
qft63	63	8	186.8	13.735537576675416	186.0	72.39948501586915	217.0	194.30791482925414
qugan71	71	6	39.4	1.1217761993408204	39.0	9.615917015075684	34.0	48.185165357589725
qugan71	71	8	84.4	2.087785339355469	84.0	8.262512731552125	44.8	66.8085223197937
swaptest83	83	6	5.0	0.8511085987091065	5.0	9.602023935317993	5.0	86.95234041213989
swaptest83	83	8	22.8	2.201212453842163	18.4	12.083254671096801	8.0	136.2230197906494
wstate76	76	6	32.4	0.23173298835754394	30.2	0.9064919948577881	8.0	18.57373332977295
wstate76	76	8	32.4	0.25309157371520996	33.8	0.9321967124938965	12.2	26.24423122406006
\end{filecontents*}
\pgfplotstabletypeset[
      col sep=space,
      header=true,
      every head row/.style={
          before row={\toprule},
          after row={\midrule},
      },
      every last row/.style={
          after row={\bottomrule},
      },
      columns={circuit_name, num_partitions, pe_cost, pe_time, esd_cost, esd_time,
               r_cost, r_time},
               columns/circuit_name/.style={
                column name={Circuit},
                column type=l,
                string type,
                string replace={_}{\_}, % escapes underscores
            },
            columns/circuit_name/.style={
        column name={Circuit},
        column type=l,
        string type,
        string replace={_}{ }, % escapes underscores
    },
    columns/num_partitions/.style={
        column name={Parts},
        column type=l,
        fixed,
        precision = 0, % escapes underscores
    },
    columns/pe_cost/.style={
        column name={PE Cost},
        column type=l,
        fixed,
        precision = 1, % escapes underscores
    },
    columns/pe_time/.style={
        column name={PE Time},
        column type=l,
        fixed,
        precision = 2, % escapes underscores
    },
    columns/esd_cost/.style={
        column name={ESD Cost},
        column type=l,
        fixed,
        precision = 1, % escapes underscores
    },
    columns/esd_time/.style={
        column name={ESD Time},
        column type=l,
        fixed,
        precision = 2, % escapes underscores
    },
    columns/r_cost/.style={
        column name={FM Cost},
        column type=l,
        fixed,
        precision = 1, % escapes underscores
    },
    columns/r_time/.style={
        column name={FM Time},
        column type=l,
        fixed,
        precision = 2, % escapes underscores
    }
    ]{PGF/QASM/merged_results_linear_nq_50.dat}
}

\end{minipage}
\hfill
\begin{minipage}{0.48\textwidth}
\centering
\centering
\caption{QASM results for direct partitioning on grid networks.}
\label{tab:QASM_res_grid}
\resizebox{\linewidth}{!}{%
    \begin{filecontents*}{PGF/QASM/merged_results_grid_nq_50.dat}
circuit_name	num_qubits	num_partitions	pe_cost	pe_time	esd_cost	esd_time	r_cost	r_time
QV100	100	6	0	0	0	0	7203.000	122.741
QV100	100	8	0	0	0	0	9470.000	170.050
adder64	64	6	26.2	1.122273063659668	24.6	6.363174104690552	17.8	36.41612620353699
adder64	64	8	58.8	1.1261861801147461	54.0	6.486672735214233	13.2	54.94883832931519
bv70	70	6	4.6	0.07433366775512695	4.6	0.13377704620361328	5.0	4.637663745880127
bv70	70	8	5.0	0.1420670509338379	5.8	0.24979982376098633	5.6	6.390420866012573
cat65	65	6	20.4	0.08715524673461914	18.0	0.1533456802368164	7.0	5.480801439285278
cat65	65	8	30.2	0.07940716743469238	30.8	0.18674616813659667	11.0	7.757790470123291
cc64	64	6	5.2	0.1852935791015625	5.0	0.3411539077758789	5.0	16.13063039779663
cc64	64	8	7.0	0.12460269927978515	7.0	0.31608967781066893	7.0	24.34614815711975
dnn51	51	6	73.2	1.7285896778106689	74.2	5.16161618232727	27.6	14.37039794921875
dnn51	51	8	108.8	2.1551465511322023	104.8	7.0798659324646	29.0	20.299654483795166
ghz78	78	6	7.0	0.06906261444091796	7.0	0.14748473167419435	7.0	8.253113412857056
ghz78	78	8	35.6	0.11661601066589355	36.0	0.2619762897491455	11.0	11.509365320205688
ising66	66	6	8.0	0.13629889488220215	8.0	0.5563277721405029	7.0	0.5796419620513916
ising66	66	8	39.6	0.1551032543182373	36.0	0.6472151756286622	11.0	0.8638099670410156
ising98	98	6	24.0	0.32935752868652346	24.2	1.5099944114685058	7.0	0.915126371383667
ising98	98	8	47.6	0.5147583484649658	44.8	1.6418017864227294	11.0	1.3074373245239257
knn67	67	6	10.2	0.9707107067108154	11.6	4.990153217315674	5.0	34.89007415771484
knn67	67	8	22.8	1.9561460494995118	19.6	7.630536794662476	8.0	56.226136207580566
qft63	63	6	103.4	16.77137384414673	103.0	82.84934148788452	150.0	100.50295195579528
qft63	63	8	139.0	12.988823509216308	139.0	69.17064504623413	217.0	147.47376537322998
qugan71	71	6	56.8	1.1106683731079101	55.0	9.313678646087647	32.2	26.748239755630493
qugan71	71	8	64.4	1.795739507675171	64.0	8.468819284439087	51.8	37.87299370765686
swaptest83	83	6	5.0	0.8127249717712403	5.0	9.668253135681152	5.0	59.54033627510071
swaptest83	83	8	26.4	2.568936014175415	28.0	13.334004020690918	8.0	94.71950273513794
wstate76	76	6	31.2	0.23181986808776855	29.2	0.9153796672821045	10.8	7.946434497833252
wstate76	76	8	48.2	0.24371004104614258	45.0	0.9881468772888183	18.2	11.272783231735229
\end{filecontents*}
\pgfplotstabletypeset[
      col sep=space,
      header=true,
      every head row/.style={
          before row={\toprule},
          after row={\midrule},
      },
      every last row/.style={
          after row={\bottomrule},
      },
      columns={circuit_name, num_partitions, pe_cost, pe_time, esd_cost, esd_time,
               r_cost, r_time},
               columns/circuit_name/.style={
                column name={Circuit},
                column type=l,
                string type,
                string replace={_}{\_}, % escapes underscores
            },
            columns/circuit_name/.style={
        column name={Circuit},
        column type=l,
        string type,
        string replace={_}{ }, % escapes underscores
    },
    columns/num_partitions/.style={
        column name={Parts},
        column type=l,
        fixed,
        precision = 0, % escapes underscores
    },
    columns/pe_cost/.style={
        column name={PE Cost},
        column type=l,
        fixed,
        precision = 1, % escapes underscores
    },
    columns/pe_time/.style={
        column name={PE Time},
        column type=l,
        fixed,
        precision = 2, % escapes underscores
    },
    columns/esd_cost/.style={
        column name={ESD Cost},
        column type=l,
        fixed,
        precision = 1, % escapes underscores
    },
    columns/esd_time/.style={
        column name={ESD Time},
        column type=l,
        fixed,
        precision = 2, % escapes underscores
    },
    columns/r_cost/.style={
        column name={FM Cost},
        column type=l,
        fixed,
        precision = 1, % escapes underscores
    },
    columns/r_time/.style={
        column name={FM Time},
        column type=l,
        fixed,
        precision = 2, % escapes underscores
    }
    ]{PGF/QASM/merged_results_grid_nq_50.dat}
}

\end{minipage}
\end{table*}

Figure \ref{fig:combined_CP} shows the performance of the heterogeneous partitioning scheme on linear, grid, and random networks using $CP$-fraction circuits, that are fixed-depth, random circuits with a varying proportion of two-qubit gates. When comparing with the Pytket-DQC methods, there are clear regions (two-qubit gate fraction $0.1-0.6$) where the temporal FM outperforms the Pytket-DQC methods for all network topologies. For high fractions of two-qubit gates, Pytket-DQC methods achieve lower entanglement costs. A possible reason for this is that high two-qubit gate density leads to control chains, thus favouring methods focused on gate teleportation. As the proportion of two-qubit gates increases, methods that focus on gate teleportation will perform better. 

\begin{figure*}[!htpb]
    \centering
    % First row: Linear topologies
    \begin{subfigure}{0.32\linewidth}
        \centering
        \usetikzlibrary{patterns}
\begin{tikzpicture}
    \begin{axis}[
        width=\columnwidth,
        height=0.65\columnwidth,
        ybar,                     % Use bars
        bar width=2.0pt,         % Control bar thickness
        symbolic x coords={4,8,16,32,64},
        unbounded coords=discard, % Discard unbound coords
        xtick={4,8,16,32,64},
        enlarge x limits=0.1,    % A bit of horizontal padding
        ymin=0, ymax=40000,      % Adjust ymax based on your data range
        ylabel={Entanglement cost},
        grid=none,
        minor grid style={line width=0.5pt, draw=gray!60, densely dotted},
        major grid style={dotted, draw=gray!70},
        % Legend positioned inside the axis
        legend style={
            at={(0.02,0.98)},
            anchor=north west,
            legend columns=1,
            font=\tiny,
            fill=white,
            fill opacity=0.8,
            draw=black,
            inner sep=3pt,
        },
    ]
    
    % Load data
    \pgfplotstableread[col sep=space,header=has colnames]{PGF/CP/NetCoarse/new_results/mean_results_128_linear.dat}\dataLinearChi
    \pgfplotstableread[col sep=space,header=has colnames]{PGF/CP/NetCoarse/new_results/direct_mean_results_128_linear.dat}\dataLinearDirect

    % Direct method using direct data file - leftmost
    \addplot+ [
        color=orange,
        fill opacity=0.4,
        bar shift=-6pt,
    ] table [
        x=num_partitions,
        y=mean_cost_final,
    ] {\dataLinearDirect};
    \addlegendentry{Direct}

    % Coarsening factor χ = 2
    \addplot+ [
        color=blue,
        fill opacity=0.3,
        bar shift=-2pt,
        restrict expr to domain={\thisrow{coarsening_factor}}{2:2},
    ] table [
        x=num_partitions,
        y=mean_cost_final,
    ] {\dataLinearChi};
    \addlegendentry{$\chi$=2}

    % Coarsening factor χ = 4
    \addplot+ [
        color=red,
        fill opacity=0.3,
        bar shift=2pt,
        restrict expr to domain={\thisrow{coarsening_factor}}{4:4},
    ] table [
        x=num_partitions,
        y=mean_cost_final,
    ] {\dataLinearChi};
    \addlegendentry{$\chi$=4}

    \addplot+ [
        color=teal,
        fill opacity=0.3,
        bar shift=-200pt,
        restrict expr to domain={\thisrow{coarsening_factor}}{4:4},
    ] table [
        x=num_partitions,
        y=mean_cost_final,
    ] {\dataLinearChi};

    % Coarsening factor χ = 8
    \addplot+ [
        color=teal,
        fill opacity=0.3,
        bar shift=6pt,
        restrict expr to domain={\thisrow{coarsening_factor}}{8:8},
    ] table [
        x=num_partitions,
        y=mean_cost_final,
    ] {\dataLinearChi};
    \addlegendentry{$\chi$=8}
    
    \end{axis}

\end{tikzpicture}
        \caption{Linear 128-qubit}
        \label{fig:linear_128}
    \end{subfigure}%
    \hfill
    \begin{subfigure}{0.32\linewidth}
        \centering
        \usetikzlibrary{patterns}
\begin{tikzpicture}
    \begin{axis}[
        width=\columnwidth,
        height=0.65\columnwidth,
        ybar,                     % Use bars
        bar width=2.0pt,         % Control bar thickness
        symbolic x coords={4,8,16,32,64},
        unbounded coords=discard, % Discard unbound coords
        xtick={4,8,16,32,64},
        enlarge x limits=0.1,    % A bit of horizontal padding
        ymin=0, ymax=190000,     % Increased ymax for 256 linear data
        grid=none,
        minor grid style={line width=0.5pt, draw=gray!60, densely dotted},
        major grid style={dotted, draw=gray!70},
        scaled y ticks = base 10:-3,  % Scale for larger values
        % Legend removed - using shared legend
    ]
    
    % Load data
    \pgfplotstableread[col sep=space,header=has colnames]{PGF/CP/NetCoarse/new_results/mean_results_256_linear.dat}\dataLinearChi
    \pgfplotstableread[col sep=space,header=has colnames]{PGF/CP/NetCoarse/new_results/direct_mean_results_256_linear.dat}\dataLinearDirect

    % Direct method using direct data file - leftmost
    \addplot+ [
        color=orange,
        fill opacity=0.4,
        bar shift=-6pt,
    ] table [
        x=num_partitions,
        y=mean_cost_final,
    ] {\dataLinearDirect};
    % \addlegendentry{Direct (NG)}

    % Coarsening factor χ = 2
    \addplot+ [
        color=blue,
        fill opacity=0.3,
        bar shift=-2pt,
        restrict expr to domain={\thisrow{coarsening_factor}}{2:2},
    ] table [
        x=num_partitions,
        y=mean_cost_final,
    ] {\dataLinearChi};
    % \addlegendentry{$\chi$=2}

    % Coarsening factor χ = 4
    \addplot+ [
        color=red,
        fill opacity=0.3,
        bar shift=2pt,
        restrict expr to domain={\thisrow{coarsening_factor}}{4:4},
    ] table [
        x=num_partitions,
        y=mean_cost_final,
    ] {\dataLinearChi};
    % \addlegendentry{$\chi$=4}

        \addplot+ [
        color=white,
        fill opacity=0.0,
        bar shift=-200pt,
        restrict expr to domain={\thisrow{coarsening_factor}}{4:4},
    ] table [
        x=num_partitions,
        y=mean_cost_final,
    ] {\dataLinearChi};

    % Coarsening factor χ = 8
    \addplot+ [
        color=teal,
        fill opacity=0.3,
        bar shift=6pt,
        restrict expr to domain={\thisrow{coarsening_factor}}{8:8},
    ] table [
        x=num_partitions,
        y=mean_cost_final,
    ] {\dataLinearChi};
    % \addlegendentry{$\chi$=8}
    
    \end{axis}
\end{tikzpicture}
        \caption{Linear 256-qubit}
        \label{fig:linear_256}
    \end{subfigure}%
    \hfill
    \begin{subfigure}{0.32\linewidth}
        \centering
        \usetikzlibrary{patterns}
\begin{tikzpicture}
    \begin{axis}[
        width=\columnwidth,
        height=0.65\columnwidth,
        ybar,                     % Use bars
        bar width=2.0pt,         % Control bar thickness
        symbolic x coords={4,8,16,32,64},
        unbounded coords=discard, % Discard unbound coords
        xtick={4,8,16,32,64},
        enlarge x limits=0.1,    % A bit of horizontal padding
        ymin=0, ymax=850000,     % Increased ymax for 512 linear data
        grid=none,
        minor grid style={line width=0.5pt, draw=gray!60, densely dotted},
        major grid style={dotted, draw=gray!70},
        scaled y ticks = base 10:-3,  % Scale for larger values
        % Legend removed - using shared legend
    ]
    
    % Load data
    \pgfplotstableread[col sep=space,header=has colnames]{PGF/CP/NetCoarse/new_results/mean_results_512_linear.dat}\dataLinearChi
    \pgfplotstableread[col sep=space,header=has colnames]{PGF/CP/NetCoarse/new_results/direct_mean_results_512_linear.dat}\dataLinearDirect
    \addplot+ [
        color=orange,
        fill opacity=0.4,
        bar shift=-6pt,
    ] table [
        x=num_partitions,
        y=mean_cost_final,
    ] {\dataLinearDirect};
    % Coarsening factor χ = 2
    \addplot+ [
        color=blue,
        fill opacity=0.3,
        bar shift=-2pt,
        restrict expr to domain={\thisrow{coarsening_factor}}{2:2},
    ] table [
        x=num_partitions,
        y=mean_cost_final,
    ] {\dataLinearChi};
    % \addlegendentry{$\chi$=2}

    % Coarsening factor χ = 4
    \addplot+ [
        color=red,
        fill opacity=0.3,
        bar shift=2pt,
        restrict expr to domain={\thisrow{coarsening_factor}}{4:4},
    ] table [
        x=num_partitions,
        y=mean_cost_final,
    ] {\dataLinearChi};
    % \addlegendentry{$\chi$=4}

    \addplot+ [
        color=teal,
        fill opacity=0.0,
        bar shift=-200pt,
        restrict expr to domain={\thisrow{coarsening_factor}}{4:4},
    ] table [
        x=num_partitions,
        y=mean_cost_final,
    ] {\dataLinearChi};

    % Coarsening factor χ = 8
    \addplot+ [
        color=teal,
        fill opacity=0.3,
        bar shift=6pt,
        restrict expr to domain={\thisrow{coarsening_factor}}{8:8},
    ] table [
        x=num_partitions,
        y=mean_cost_final,
    ] {\dataLinearChi};
    % \addlegendentry{$\chi$=8}
    
    \end{axis}
\end{tikzpicture}
        \caption{Linear 512-qubit}
        \label{fig:linear_512}
    \end{subfigure}%
    
    \vspace{0.1cm}
    
    % Second row: Grid topologies
    \begin{subfigure}{0.32\linewidth}
        \centering
        \usetikzlibrary{patterns}
\begin{tikzpicture}
    \begin{axis}[
        width=\columnwidth,
        height=0.65\columnwidth,
        ybar,                     % Use bars
        bar width=2.0pt,         % Control bar thickness
        symbolic x coords={4,8,16,32,64},
        unbounded coords=discard, % Discard unbound coords
        xtick={4,8,16,32,64},
        enlarge x limits=0.1,    % A bit of horizontal padding
        ymin=0, ymax=15500,      % Adjust ymax based on your data range
        ylabel={Entanglement cost},
        grid=none,
        minor grid style={line width=0.5pt, draw=gray!60, densely dotted},
        major grid style={dotted, draw=gray!70},
        % Legend removed - using shared legend
    ]
    
    % Load data
    \pgfplotstableread[col sep=space,header=has colnames]{PGF/CP/NetCoarse/new_results/mean_results_128_grid.dat}\dataGridChi
    \pgfplotstableread[col sep=space,header=has colnames]{PGF/CP/NetCoarse/new_results/direct_mean_results_128_grid.dat}\dataGridDirect

    % Direct method using direct data file - leftmost
    \addplot+ [
        color=orange,
        fill opacity=0.4,
        bar shift=-6pt,
    ] table [
        x=num_partitions,
        y=mean_cost_final,
    ] {\dataGridDirect};
    % \addlegendentry{Direct (NG)}

    % Coarsening factor χ = 2
    \addplot+ [
        color=blue,
        fill opacity=0.3,
        bar shift=-2pt,
        restrict expr to domain={\thisrow{coarsening_factor}}{2:2},
    ] table [
        x=num_partitions,
        y=mean_cost_final,
    ] {\dataGridChi};
    % \addlegendentry{$\chi$=2}

    % Coarsening factor χ = 4
    \addplot+ [
        color=red,
        fill opacity=0.3,
        bar shift=2pt,
        restrict expr to domain={\thisrow{coarsening_factor}}{4:4},
    ] table [
        x=num_partitions,
        y=mean_cost_final,
    ] {\dataGridChi};
    % \addlegendentry{$\chi$=4}

        \addplot+ [
        color=teal,
        fill opacity=0.0,
        bar shift=-200pt,
        restrict expr to domain={\thisrow{coarsening_factor}}{4:4},
    ] table [
        x=num_partitions,
        y=mean_cost_final,
    ] {\dataGridChi};

    % Coarsening factor χ = 8
    \addplot+ [
        color=teal,
        fill opacity=0.3,
        bar shift=6pt,
        restrict expr to domain={\thisrow{coarsening_factor}}{8:8},
    ] table [
        x=num_partitions,
        y=mean_cost_final,
    ] {\dataGridChi};
    % \addlegendentry{$\chi$=8}
    
    \end{axis}
\end{tikzpicture}
        \caption{Grid 128-qubit}
        \label{fig:grid_128}
    \end{subfigure}%
    \hfill
    \begin{subfigure}{0.32\linewidth}
        \centering
        \usetikzlibrary{patterns}
\begin{tikzpicture}
    \begin{axis}[
        width=\columnwidth,
        height=0.65\columnwidth,
        ybar,                     % Use bars
        bar width=2.0pt,         % Control bar thickness
        symbolic x coords={4,8,16,32,64},
        unbounded coords=discard, % Discard unbound coords
        xtick={4,8,16,32,64},
        enlarge x limits=0.1,    % A bit of horizontal padding
        ymin=0, ymax=67000,      % Adjust ymax based on your data range
        grid=none,
        minor grid style={line width=0.5pt, draw=gray!60, densely dotted},
        major grid style={dotted, draw=gray!70},
        % Legend removed - using shared legend
    ]
    
    % Load data
    \pgfplotstableread[col sep=space,header=has colnames]{PGF/CP/NetCoarse/new_results/mean_results_256_grid.dat}\dataGridChi
    \pgfplotstableread[col sep=space,header=has colnames]{PGF/CP/NetCoarse/new_results/direct_mean_results_256_grid.dat}\dataGridDirect
    
        \addplot+ [
        color=orange,
        fill opacity=0.4,
        bar shift=-6pt,
    ] table [
        x=num_partitions,
        y=mean_cost_final,
    ] {\dataGridDirect};

    % Coarsening factor χ = 2
    \addplot+ [
        color=blue,
        fill opacity=0.3,
        bar shift=-2pt,
        restrict expr to domain={\thisrow{coarsening_factor}}{2:2},
    ] table [
        x=num_partitions,
        y=mean_cost_final,
    ] {\dataGridChi};
    % \addlegendentry{$\chi$=2}

    % Coarsening factor χ = 4
    \addplot+ [
        color=red,
        fill opacity=0.3,
        bar shift=2pt,
        restrict expr to domain={\thisrow{coarsening_factor}}{4:4},
    ] table [
        x=num_partitions,
        y=mean_cost_final,
    ] {\dataGridChi};
    % \addlegendentry{$\chi$=4}

    \addplot+ [
        color=teal,
        fill opacity=0.0,
        bar shift=-200pt,
        restrict expr to domain={\thisrow{coarsening_factor}}{4:4},
    ] table [
        x=num_partitions,
        y=mean_cost_final,
    ] {\dataGridChi};

    % Coarsening factor χ = 8
    \addplot+ [
        color=teal,
        fill opacity=0.3,
        bar shift=6pt,
        restrict expr to domain={\thisrow{coarsening_factor}}{8:8},
    ] table [
        x=num_partitions,
        y=mean_cost_final,
    ] {\dataGridChi};
    % \addlegendentry{$\chi$=8}
    
    \end{axis}
\end{tikzpicture}
        \caption{Grid 256-qubit}
        \label{fig:grid_256}
    \end{subfigure}%
    \hfill
    \begin{subfigure}{0.32\linewidth}
        \centering
        \usetikzlibrary{patterns}
\begin{tikzpicture}
    \begin{axis}[
        width=\columnwidth,
        height=0.65\columnwidth,
        ybar,                     % Use bars
        bar width=2.0pt,         % Control bar thickness
        symbolic x coords={4,8,16,32,64},
        unbounded coords=discard, % Discard unbound coords
        xtick={4,8,16,32,64},
        enlarge x limits=0.1,    % A bit of horizontal padding
        ymin=0, ymax=300000,     % Increased ymax for 512 linear data
        grid=none,
        minor grid style={line width=0.5pt, draw=gray!60, densely dotted},
        major grid style={dotted, draw=gray!70},
        scaled y ticks = base 10:-3,  % Scale for larger values
        % Legend removed - using shared legend
    ]
    
    % Load data
    \pgfplotstableread[col sep=space,header=has colnames]{PGF/CP/NetCoarse/new_results/mean_results_512_grid.dat}\dataGridChi
    \pgfplotstableread[col sep=space,header=has colnames]{PGF/CP/NetCoarse/new_results/direct_mean_results_512_grid.dat}\dataGridDirect
    \addplot+ [
        color=orange,
        fill opacity=0.4,
        bar shift=-6pt,
    ] table [
        x=num_partitions,
        y=mean_cost_final,
    ] {\dataGridDirect};
    % Coarsening factor χ = 2
    \addplot+ [
        color=blue,
        fill opacity=0.3,
        bar shift=-2pt,
        restrict expr to domain={\thisrow{coarsening_factor}}{2:2},
    ] table [
        x=num_partitions,
        y=mean_cost_final,
    ] {\dataGridChi};
    % \addlegendentry{$\chi$=2}

    % Coarsening factor χ = 4
    \addplot+ [
        color=red,
        fill opacity=0.3,
        bar shift=2pt,
        restrict expr to domain={\thisrow{coarsening_factor}}{4:4},
    ] table [
        x=num_partitions,
        y=mean_cost_final,
    ] {\dataGridChi};
    % \addlegendentry{$\chi$=4}

    \addplot+ [
        color=teal,
        fill opacity=0.0,
        bar shift=-200pt,
        restrict expr to domain={\thisrow{coarsening_factor}}{4:4},
    ] table [
        x=num_partitions,
        y=mean_cost_final,
    ] {\dataGridChi};

    % Coarsening factor χ = 8
    \addplot+ [
        color=teal,
        fill opacity=0.3,
        bar shift=6pt,
        restrict expr to domain={\thisrow{coarsening_factor}}{8:8},
    ] table [
        x=num_partitions,
        y=mean_cost_final,
    ] {\dataGridChi};
    % \addlegendentry{$\chi$=8}
    
    \end{axis}
\end{tikzpicture}
        \caption{Grid 512-qubit}
        \label{fig:grid_512}
    \end{subfigure}%
 
    \centering
    % First row: Linear topologies (time)
    \begin{subfigure}{0.32\linewidth}
        \centering
        \usetikzlibrary{patterns}
\begin{tikzpicture}
    \begin{axis}[
        width=\columnwidth,
        height=0.65\columnwidth,
        ybar,                     % Use bars
        bar width=2.0pt,         % Control bar thickness
        symbolic x coords={4,8,16,32,64},
        unbounded coords=discard, % Discard unbound coords
        xtick={4,8,16,32,64},
        enlarge x limits=0.1,    % A bit of horizontal padding
        ymin=0, ymax=140,         % Adjust ymax based on time data range
        ylabel={Time taken (s)},
        scaled y ticks=false,
        scaled y ticks = base 10:-2,  % Scale for larger values
        grid=none,
        minor grid style={line width=0.5pt, draw=gray!60, densely dotted},
        major grid style={dotted, draw=gray!70},
        % Legend positioned inside the axis
        legend style={
            at={(0.02,0.98)},
            anchor=north west,
            legend columns=1,
            font=\tiny,
            fill=white,
            fill opacity=0.8,
            draw=black,
            inner sep=3pt,
        },
    ]
    
    % Load data
    \pgfplotstableread[col sep=space,header=has colnames]{PGF/CP/NetCoarse/new_results/mean_results_128_linear.dat}\dataLinearChi
    \pgfplotstableread[col sep=space,header=has colnames]{PGF/CP/NetCoarse/new_results/direct_mean_results_128_linear.dat}\dataLinearDirect

    % Direct method using direct data file - leftmost
    \addplot+ [
        color=orange,
        fill opacity=0.4,
        bar shift=-6pt,
    ] table [
        x=num_partitions,
        y=mean_time_taken,
    ] {\dataLinearDirect};
    % \addlegendentry{Direct}

    % Coarsening factor χ = 2
    \addplot+ [
        color=blue,
        fill opacity=0.3,
        bar shift=-2pt,
        restrict expr to domain={\thisrow{coarsening_factor}}{2:2},
    ] table [
        x=num_partitions,
        y=mean_time_taken,
    ] {\dataLinearChi};
    % \addlegendentry{$\chi$=2}

    % Coarsening factor χ = 4
    \addplot+ [
        color=red,
        fill opacity=0.3,
        bar shift=2pt,
        restrict expr to domain={\thisrow{coarsening_factor}}{4:4},
    ] table [
        x=num_partitions,
        y=mean_time_taken,
    ] {\dataLinearChi};
    % \addlegendentry{$\chi$=4}

        \addplot+ [
        color=teal,
        fill opacity=0.3,
        bar shift=-200pt,
        restrict expr to domain={\thisrow{coarsening_factor}}{4:4},
    ] table [
        x=num_partitions,
        y=mean_time_taken,
    ] {\dataLinearChi};

    % Coarsening factor χ = 8
    \addplot+ [
        color=teal,
        fill opacity=0.3,
        bar shift=6pt,
        restrict expr to domain={\thisrow{coarsening_factor}}{8:8},
    ] table [
        x=num_partitions,
        y=mean_time_taken,
    ] {\dataLinearChi};
    % \addlegendentry{$\chi$=8}
    
    \end{axis}
\end{tikzpicture}
        \caption{Linear 128-qubit}
        \label{fig:linear_128_time}
    \end{subfigure}%
    \hfill
    \begin{subfigure}{0.32\linewidth}
        \centering
        \usetikzlibrary{patterns}
\begin{tikzpicture}
    \begin{axis}[
        width=\columnwidth,
        height=0.65\columnwidth,
        ybar,                     % Use bars
        bar width=2.0pt,         % Control bar thickness
        symbolic x coords={4,8,16,32,64},
        unbounded coords=discard, % Discard unbound coords
        xtick={4,8,16,32,64},
        enlarge x limits=0.1,    % A bit of horizontal padding
        ymin=0, ymax=1100,        % Adjusted for 256 time data range
        scaled y ticks=false,
        scaled y ticks = base 10:-2,  % Scale for larger values
        grid=none,
        minor grid style={line width=0.5pt, draw=gray!60, densely dotted},
        major grid style={dotted, draw=gray!70},
        % Legend positioned inside the axis
        legend style={
            at={(0.02,0.98)},
            anchor=north west,
            legend columns=1,
            font=\tiny,
            fill=white,
            fill opacity=0.8,
            draw=black,
            inner sep=3pt,
        },
    ]
    
    % Load data
    \pgfplotstableread[col sep=space,header=has colnames]{PGF/CP/NetCoarse/new_results/mean_results_256_linear.dat}\dataLinearChi
    \pgfplotstableread[col sep=space,header=has colnames]{PGF/CP/NetCoarse/new_results/direct_mean_results_256_linear.dat}\dataLinearDirect

    % Direct method using direct data file - leftmost
    \addplot+ [
        color=orange,
        fill opacity=0.4,
        bar shift=-6pt,
    ] table [
        x=num_partitions,
        y=mean_time_taken,
    ] {\dataLinearDirect};
    % \addlegendentry{Direct}

    % Coarsening factor χ = 2
    \addplot+ [
        color=blue,
        fill opacity=0.3,
        bar shift=-2pt,
        restrict expr to domain={\thisrow{coarsening_factor}}{2:2},
    ] table [
        x=num_partitions,
        y=mean_time_taken,
    ] {\dataLinearChi};
    % \addlegendentry{$\chi$=2}

    % Coarsening factor χ = 4
    \addplot+ [
        color=red,
        fill opacity=0.3,
        bar shift=2pt,
        restrict expr to domain={\thisrow{coarsening_factor}}{4:4},
    ] table [
        x=num_partitions,
        y=mean_time_taken,
    ] {\dataLinearChi};
    % \addlegendentry{$\chi$=4}

    \addplot+ [
        color=teal,
        fill opacity=0.0,
        bar shift=-200pt,
        restrict expr to domain={\thisrow{coarsening_factor}}{4:4},
    ] table [
        x=num_partitions,
        y=mean_time_taken,
    ] {\dataLinearChi};

    % Coarsening factor χ = 8
    \addplot+ [
        color=teal,
        fill opacity=0.3,
        bar shift=6pt,
        restrict expr to domain={\thisrow{coarsening_factor}}{8:8},
    ] table [
        x=num_partitions,
        y=mean_time_taken,
    ] {\dataLinearChi};
    % \addlegendentry{$\chi$=8}

    \end{axis}
\end{tikzpicture}
        \caption{Linear 256-qubit}
        \label{fig:linear_256_time}
    \end{subfigure}%
    \hfill
    \begin{subfigure}{0.32\linewidth}
        \centering
        \usetikzlibrary{patterns}
\begin{tikzpicture}
    \begin{axis}[
        width=\columnwidth,
        height=0.65\columnwidth,
        ybar,                     % Use bars
        bar width=2.0pt,         % Control bar thickness
        symbolic x coords={4,8,16,32,64},
        unbounded coords=discard, % Discard unbound coords
        xtick={4,8,16,32,64},
        enlarge x limits=0.1,    % A bit of horizontal padding
        ymin=0, ymax=8100,       % Adjusted for 512 time data range
        scaled y ticks = base 10:-2,  % Scale for larger values
        grid=none,
        minor grid style={line width=0.5pt, draw=gray!60, densely dotted},
        major grid style={dotted, draw=gray!70},
        % Legend positioned inside the axis
        legend style={
            at={(0.02,0.98)},
            anchor=north west,
            legend columns=1,
            font=\tiny,
            fill=white,
            fill opacity=0.8,
            draw=black,
            inner sep=3pt,
        },
    ]
    
    % Load data
    \pgfplotstableread[col sep=space,header=has colnames]{PGF/CP/NetCoarse/new_results/mean_results_512_linear.dat}\dataLinearChi
    \pgfplotstableread[col sep=space,header=has colnames]{PGF/CP/NetCoarse/new_results/direct_mean_results_512_linear.dat}\dataLinearDirect

    \addplot+ [
        color=orange,
        fill opacity=0.4,
        bar shift=-6pt,
    ] table [
        x=num_partitions,
        y=mean_time_taken,
    ] {\dataLinearDirect};
    % Coarsening factor χ = 2
    \addplot+ [
        color=blue,
        fill opacity=0.3,
        bar shift=-2pt,
        restrict expr to domain={\thisrow{coarsening_factor}}{2:2},
    ] table [
        x=num_partitions,
        y=mean_time_taken,
    ] {\dataLinearChi};
    % \addlegendentry{$\chi$=2}

    % Coarsening factor χ = 4
    \addplot+ [
        color=red,
        fill opacity=0.3,
        bar shift=2pt,
        restrict expr to domain={\thisrow{coarsening_factor}}{4:4},
    ] table [
        x=num_partitions,
        y=mean_time_taken,
    ] {\dataLinearChi};
    % \addlegendentry{$\chi$=4}

    \addplot+ [
        color=teal,
        fill opacity=0.0,
        bar shift=-200pt,
        restrict expr to domain={\thisrow{coarsening_factor}}{4:4},
    ] table [
        x=num_partitions,
        y=mean_time_taken,
    ] {\dataLinearChi};

    % Coarsening factor χ = 8
    \addplot+ [
        color=teal,
        fill opacity=0.3,
        bar shift=6pt,
        restrict expr to domain={\thisrow{coarsening_factor}}{8:8},
    ] table [
        x=num_partitions,
        y=mean_time_taken,
    ] {\dataLinearChi};
    % \addlegendentry{$\chi$=8}

    \end{axis}
\end{tikzpicture}
        \caption{Linear 512-qubit}
        \label{fig:linear_512_time}
    \end{subfigure}%
    
    \vspace{0.1cm}
    
    % Second row: Grid topologies (time)
    \begin{subfigure}{0.32\linewidth}
        \centering
        \usetikzlibrary{patterns}
\begin{tikzpicture}
    \begin{axis}[
        width=\columnwidth,
        height=0.65\columnwidth,
        ybar,                     % Use bars
        bar width=2.0pt,         % Control bar thickness
        symbolic x coords={4,8,16,32,64},
        unbounded coords=discard, % Discard unbound coords
        xtick={4,8,16,32,64},
        enlarge x limits=0.1,    % A bit of horizontal padding
        ymin=0, ymax=128,         % Adjust ymax based on time data range
        xlabel={Number of QPUs},
        ylabel={Time taken (s)},
        scaled y ticks=false,
        scaled y ticks = base 10:-2,  % Scale for larger values
        grid=none,
        minor grid style={line width=0.5pt, draw=gray!60, densely dotted},
        major grid style={dotted, draw=gray!70},
        % Legend removed - using shared legend
    ]
    
    % Load data
    \pgfplotstableread[col sep=space,header=has colnames]{PGF/CP/NetCoarse/new_results/mean_results_128_grid.dat}\dataGridChi
    \pgfplotstableread[col sep=space,header=has colnames]{PGF/CP/NetCoarse/new_results/direct_mean_results_128_grid.dat}\dataGridDirect

    % Placeholder for Direct method - invisible
    \addplot+ [
        color=orange,
        fill opacity=0.4,
        bar shift=-6pt,
    ] table [
        x=num_partitions,
        y=mean_time_taken,
    ] {\dataGridDirect};

    % Coarsening factor χ = 2
    \addplot+ [
        color=blue,
        fill opacity=0.3,
        bar shift=-2pt,
        restrict expr to domain={\thisrow{coarsening_factor}}{2:2},
    ] table [
        x=num_partitions,
        y=mean_time_taken,
    ] {\dataGridChi};
    % \addlegendentry{$\chi$=2}

    % Coarsening factor χ = 4
    \addplot+ [
        color=red,
        fill opacity=0.3,
        bar shift=2pt,
        restrict expr to domain={\thisrow{coarsening_factor}}{4:4},
    ] table [
        x=num_partitions,
        y=mean_time_taken,
    ] {\dataGridChi};
    % \addlegendentry{$\chi$=4}

        \addplot+ [
        color=teal,
        fill opacity=0.3,
        bar shift=-200pt,
        restrict expr to domain={\thisrow{coarsening_factor}}{4:4},
    ] table [
        x=num_partitions,
        y=mean_time_taken,
    ] {\dataGridChi};

    % Coarsening factor χ = 8
    \addplot+ [
        color=teal,
        fill opacity=0.3,
        bar shift=6pt,
        restrict expr to domain={\thisrow{coarsening_factor}}{8:8},
    ] table [
        x=num_partitions,
        y=mean_time_taken,
    ] {\dataGridChi};
    % \addlegendentry{$\chi$=8}
    
    \end{axis}
\end{tikzpicture}
        \caption{Grid 128-qubit}
        \label{fig:grid_128_time}
    \end{subfigure}%
    \hfill
    \begin{subfigure}{0.32\linewidth}
        \centering
        \usetikzlibrary{patterns}
\begin{tikzpicture}
    \begin{axis}[
        width=\columnwidth,
        height=0.65\columnwidth,
        ybar,                     % Use bars
        bar width=2.0pt,         % Control bar thickness
        symbolic x coords={4,8,16,32,64},
        unbounded coords=discard, % Discard unbound coords
        xtick={4,8,16,32,64},
        enlarge x limits=0.1,    % A bit of horizontal padding
        ymin=0, ymax=550,        % Adjusted for 256 grid time data range
        xlabel={Number of QPUs},
        scaled y ticks=false,
        scaled y ticks = base 10:-2,  % Scale for larger values
        grid=none,
        minor grid style={line width=0.5pt, draw=gray!60, densely dotted},
        major grid style={dotted, draw=gray!70},
        % Legend positioned inside the axis
        legend style={
            at={(0.02,0.98)},
            anchor=north west,
            legend columns=1,
            font=\tiny,
            fill=white,
            fill opacity=0.8,
            draw=black,
            inner sep=3pt,
        },
    ]
    
    % Load data
    \pgfplotstableread[col sep=space,header=has colnames]{PGF/CP/NetCoarse/new_results/mean_results_256_grid.dat}\dataGridChi
    \pgfplotstableread[col sep=space,header=has colnames]{PGF/CP/NetCoarse/new_results/direct_mean_results_256_grid.dat}\dataGridDirect

        \addplot+ [
        color=orange,
        fill opacity=0.4,
        bar shift=-6pt,
    ] table [
        x=num_partitions,
        y=mean_time_taken,
    ] {\dataGridDirect};

    % Coarsening factor χ = 2
    \addplot+ [
        color=blue,
        fill opacity=0.3,
        bar shift=-2pt,
        restrict expr to domain={\thisrow{coarsening_factor}}{2:2},
    ] table [
        x=num_partitions,
        y=mean_time_taken,
    ] {\dataGridChi};
    % \addlegendentry{$\chi$=2}

    % Coarsening factor χ = 4
    \addplot+ [
        color=red,
        fill opacity=0.3,
        bar shift=2pt,
        restrict expr to domain={\thisrow{coarsening_factor}}{4:4},
    ] table [
        x=num_partitions,
        y=mean_time_taken,
    ] {\dataGridChi};
    % \addlegendentry{$\chi$=4}

    \addplot+ [
        color=teal,
        fill opacity=0.0,
        bar shift=-200pt,
        restrict expr to domain={\thisrow{coarsening_factor}}{4:4},
    ] table [
        x=num_partitions,
        y=mean_time_taken,
    ] {\dataGridChi};

    % Coarsening factor χ = 8
    \addplot+ [
        color=teal,
        fill opacity=0.3,
        bar shift=6pt,
        restrict expr to domain={\thisrow{coarsening_factor}}{8:8},
    ] table [
        x=num_partitions,
        y=mean_time_taken,
    ] {\dataGridChi};
    % \addlegendentry{$\chi$=8}

    \end{axis}
\end{tikzpicture}
        \caption{Grid 256-qubit}
        \label{fig:grid_256_time}
    \end{subfigure}%
    \hfill
    \begin{subfigure}{0.32\linewidth}
        \centering
        \usetikzlibrary{patterns}
\begin{tikzpicture}
    \begin{axis}[
        width=\columnwidth,
        height=0.65\columnwidth,
        ybar,                     % Use bars
        bar width=2.0pt,         % Control bar thickness
        symbolic x coords={4,8,16,32,64},
        unbounded coords=discard, % Discard unbound coords
        xtick={4,8,16,32,64},
        enlarge x limits=0.1,    % A bit of horizontal padding
        ymin=0, ymax=5000,       % Adjusted for 512 time data range
        xlabel={Number of QPUs},
        scaled y ticks=false,
        scaled y ticks = base 10:-2,  % Scale for larger values
        grid=none,
        minor grid style={line width=0.5pt, draw=gray!60, densely dotted},
        major grid style={dotted, draw=gray!70},
        % Legend positioned inside the axis
        legend style={
            at={(0.02,0.98)},
            anchor=north west,
            legend columns=1,
            font=\tiny,
            fill=white,
            fill opacity=0.8,
            draw=black,
            inner sep=3pt,
        },
    ]
    
    % Load data
    \pgfplotstableread[col sep=space,header=has colnames]{PGF/CP/NetCoarse/new_results/mean_results_512_grid.dat}\dataGridChi
    \pgfplotstableread[col sep=space,header=has colnames]{PGF/CP/NetCoarse/new_results/direct_mean_results_512_grid.dat}\dataGridDirect

    \addplot+ [
        color=orange,
        fill opacity=0.4,
        bar shift=-6pt,
    ] table [
        x=num_partitions,
        y=mean_time_taken,
    ] {\dataGridDirect};
    % \addlegendentry{Direct}

    % Coarsening factor χ = 2
    \addplot+ [
        color=blue,
        fill opacity=0.3,
        bar shift=-2pt,
        restrict expr to domain={\thisrow{coarsening_factor}}{2:2},
    ] table [
        x=num_partitions,
        y=mean_time_taken,
    ] {\dataGridChi};
    % \addlegendentry{$\chi$=2}

    % Coarsening factor χ = 4
    \addplot+ [
        color=red,
        fill opacity=0.3,
        bar shift=2pt,
        restrict expr to domain={\thisrow{coarsening_factor}}{4:4},
    ] table [
        x=num_partitions,
        y=mean_time_taken,
    ] {\dataGridChi};
    % \addlegendentry{$\chi$=4}

        \addplot+ [
            color=teal,
            fill opacity=0.0,
        bar shift=-200pt,
        restrict expr to domain={\thisrow{coarsening_factor}}{4:4},
    ] table [
        x=num_partitions,
        y=mean_time_taken,
    ] {\dataGridChi};

    % Coarsening factor χ = 8
    \addplot+ [
        color=teal,
        fill opacity=0.3,
        bar shift=6pt,
        restrict expr to domain={\thisrow{coarsening_factor}}{8:8},
    ] table [
        x=num_partitions,
        y=mean_time_taken,
    ] {\dataGridChi};
    % \addlegendentry{$\chi$=8}

    \end{axis}
\end{tikzpicture}
        \caption{Grid 512-qubit}
        \label{fig:grid_512_time}
    \end{subfigure}%

    \caption{Entanglement costs and runtime for partitioning with network coarsening. For smaller networks we use coarsening factors $\chi=2$ and $\chi=4$, while for larger networks we use $\chi=2$, $\chi=4$ and $\chi=8$. Results show mean entanglement costs and runtimes across different numbers of QPUs for linear and grid topologies. Direct partitioning results are shown for comparison below 16 QPUs, after which an unoptimised layout must be used.}
    \label{fig:coarsening_results}

    % \caption{Runtime performance for different network topologies and circuit sizes showing the effect of network coarsening on partitioning time. For larger instances direct partitioning is not feasible, and so we only show the network coarsening results.}
    % \label{fig:coarsening_time_results}
\end{figure*}
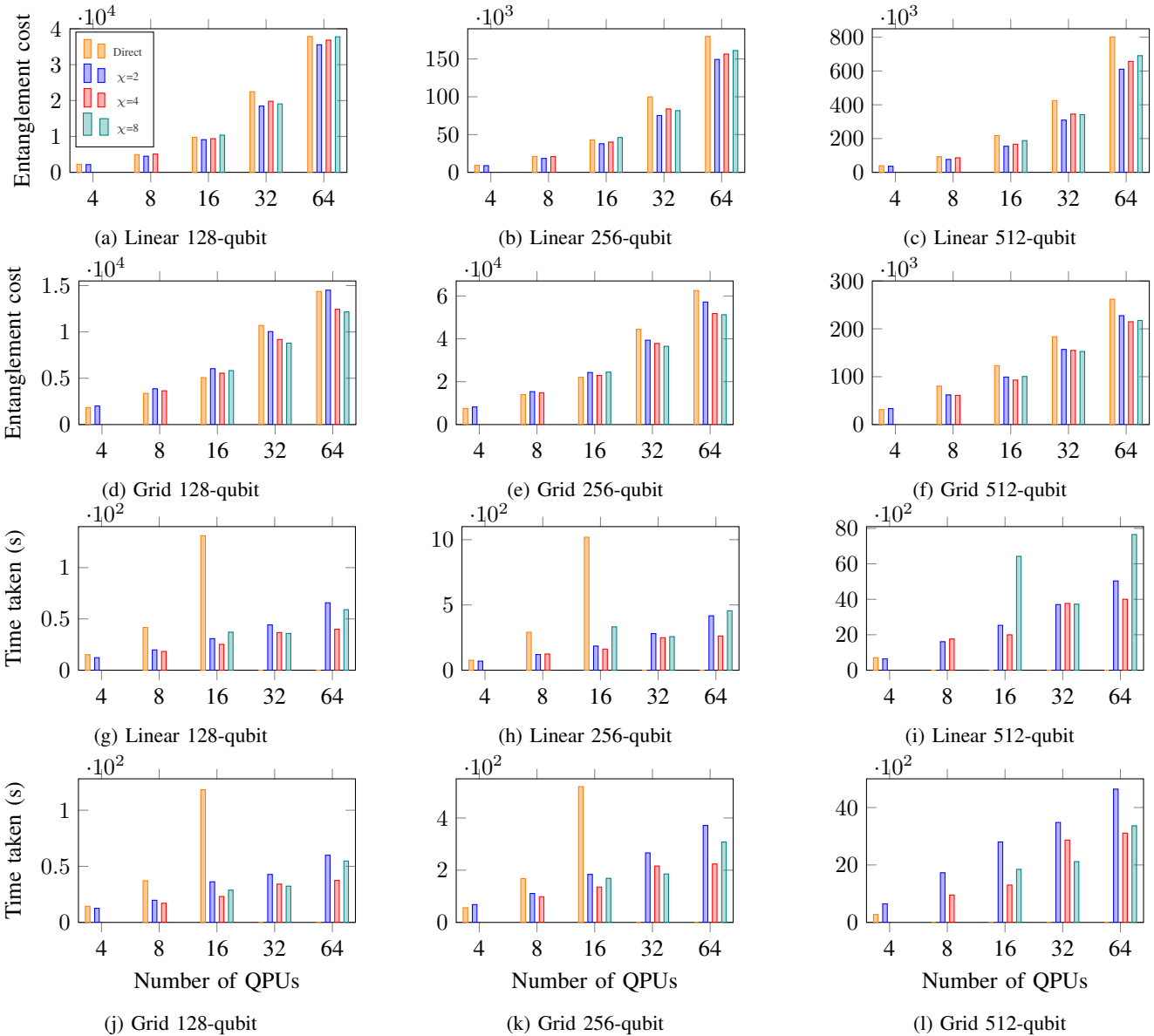

For real circuits, we use the QASM benchmark suite, comparing results with the PartitionEmbed and EmbedSteinerDetached methods from Pytket-DQC, two methods that are the best performers from Andres-Martinez et al. \cite{andres-martinez_distributing_2024}. These results are shown for linear topologies of $6$ and $8$ QPUs in Tab. \ref{tab:QASM_res_linear}, and for grid topologies in Tab. \ref{tab:QASM_res_grid}. It can be seen from these results that our methods are effective in terms of entanglement costs, outperforming both baseline methods in almost all cases. However, we find the direct partitioning to be slower in most cases. It is for this reason that we require network coarsening for larger problem sizes.

\subsection{Large-scale networks}\label{sec:large-scale-res}

We assess the performance of the network coarsening, using 128-, 256- and 512-qubit square $CP$-fraction circuits, for linear and grid topologies. We create linear and grid networks ranging from $4$ to $64$ QPUs, and coarsen them using factors $\chi=2$, $\chi=4$ and $\chi=8$ where applicable. The entanglement cost and runtime results are shown in Fig. \ref{fig:coarsening_results}. Network coarsening achieves significant speedup, while maintaining similar or improved solution quality. An immediate advantage is achieved by the coarsened methods as soon as direct partitioning is not feasible.

\section{Discussion and Conclusions}\label{sec:conclusion}
The results from Sec. \ref{sec:intermediate-scale-res} show that the generalisation of temporally coarsened FM to arbitrary networks is effective, achieving lower entanglement costs than state-of-the-art methods in most cases. This is demonstrated for random circuits (Fig. \ref{fig:combined_CP}), as well as well-known circuits from the QASM benchmark suite (Tab. \ref{tab:QASM_res_linear} and \ref{tab:QASM_res_grid}). In many of these results, however, we find that the run-time of direct partitioning is longer than the baseline methods. When evaluating the network coarsening scheme, we find that we are able to achieve a similar or improved solution quality compared with the direct methods in most cases, while significantly reducing the run-time for large numbers of QPUs. The benefits are most clear for linear networks, where we achieve lower entanglement costs than direct partitioning in all cases. For grid networks, coarsening does not always lead to a better result than direct partitioning, particularly for smaller numbers of qubits. We interpret this difference by noting that the coarsened approach forces qubits to be more widely spread across the network, which proves effective for linear networks with many long-range links. For grid networks, which have higher connectivity, lower costs are often achieved by filling up nearby QPUs as much as possible, rather than spreading them across the network. Where the problem sizes are too large for direct partitioning, we are forced to use the cost of the initial layout as a baseline, which is optimised only in terms of the groupings for gate teleportation but not in terms of the qubit assignment. This results in an immediate advantage for coarsened method. While the run-times for large circuits and networks are still relatively high, the internal multilevel partitioning can be capped at a lower level to further reduce run-times.
Overall, the results indicate that network coarsening is a promising approach for optimising partitioning over large-scale quantum networks, and can be used to achieve good solutions in a reasonable time. We have shown that our methods for temporal partitioning of quantum circuits can be effectively extended to consider general quantum network topologies, thus directly targeting auxiliary entanglement costs. We showed that we are able to outperform state-of-the-art methods in terms of entanglement costs, albeit with longer run-time. Through exploring network coarsening we improved the efficiency and found that we were able to still match or outperform direct partitioning methods for larger networks.

\section{Future work}
We are yet to investigate network coarsening for irregular networks and other quantum circuits. There is room to explore different techniques for network coarsening. We plan to investigate these in future work, and integrate all techniques into the \texttt{disqco} library \cite{disqco2025}, an ongoing project for implementing circuit optimisation techniques for distributed architectures.

\section*{Acknowledgements}

The authors acknowledge funding from the Engineering and Physical Sciences Research Council (EPSRC), grant number EP/W032643/1.

\bibliographystyle{IEEEtran}
\bibliography{filtered_references}

\end{document}